\documentclass[12pt]{article}

\usepackage[english]{babel}
\usepackage{todonotes}
\usepackage{graphicx}
\usepackage{subcaption}
\usepackage{multirow}
\usepackage{booktabs}
\usepackage[letterpaper,top=2cm,bottom=2cm,left=3cm,right=3cm,marginparwidth=1.75cm]{geometry}

\usepackage{amsmath}
\usepackage{graphicx}
\usepackage[colorlinks=true, allcolors=blue]{hyperref}

\title{Machine-learning surrogate models for nonlinear energetic-particle transport predictions in ITER}
\author{
Yashika Ghai$^{1}$,
Donald A. Spong$^{1}$,
Jacobo Varela$^{2}$,
Luis Garcia$^{3}$\\[0.5em]
{\small $^{1}$Oak Ridge National Laboratory, Oak Ridge, TN 37831, USA}\\
{\small $^{2}$Institute for Fusion Studies, The University of Texas at Austin, Austin, TX 78712, USA}\\
{\small $^{3}$Universidad Carlos III de Madrid, Madrid, Spain}
}
\date{}
\begin{document}
\maketitle

\begin{abstract}
Fast and accurate prediction of energetic-particle transport driven by Alfvén eigenmode (AE) instabilities is essential for integrated modeling workflows used in the design and optimization of burning plasma fusion reactors. In this work, we develop machine-learning-based surrogate models for rapid prediction of energetic beam and alpha-particle transport fluxes, together with predictive uncertainty estimates, for an ITER steady-state scenario. Two complementary surrogate methodologies, Gaussian process (GP) regression and hierarchical neural networks (NNs), are trained using nonlinear FAR3d gyrofluid simulations of energetic-particle transport. A flux-variability analysis demonstrates that the selected plasma-state representation provides a sufficiently unique parameterization of the nonlinear transport response over most of the sampled feature space, thereby justifying the surrogate formulation. Both surrogate models reproduce the nonlinear transport fluxes with high predictive accuracy while reducing the computational cost of transport evaluation by approximately five to six orders of magnitude relative to direct nonlinear FAR3d simulations. Although the two approaches achieve comparable predictive accuracy, they exhibit distinct uncertainty characteristics: the GP provides more consistent global uncertainty estimates, whereas the NN more clearly distinguishes between different transport regimes. This work establishes a proof of concept for developing machine-learning surrogate models of energetic-particle transport that are sufficiently accurate and computationally efficient to be incorporated into future integrated modeling workflows.
\end{abstract}
\footnote{
Notice: This manuscript has been authored by UT-Battelle, LLC, under contract DE-AC05-00OR22725 with the US Department of Energy (DOE). The US government retains and the publisher, by accepting the article for publication, acknowledges that the US government retains a nonexclusive, paid-up, irrevocable, worldwide license to publish or reproduce the published form of this manuscript, or allow others to do so, for US government purposes. DOE will provide public access to these results of federally sponsored research in accordance with the DOE Public Access Plan (https://www.energy.gov/doe-public-access-plan).}
\section{Introduction}


\par
Burning plasma conditions are realized when the plasma is predominantly self-heated due to collisional slowing-down of fusion-born high energy (3.5 MeV) alpha particles. The Q-factor, described as the ratio of power produced by fusion reactions to the external power used to heat the plasma, is expected to be in the range of 20-50 for future fusion reactors operating at "burning plasma" conditions. To achieve Q values in this range, approximately $80-90\%$ of the born alpha particle population needs to be confined \cite{Mayoral_2004, Salewski_2025}. ITER is aimed to operate at $Q\geq10$ for a duration of about 300-500s which requires confinement of about two-third of the alpha particle population. However, energetic particle instabilities that occur at Alfvén time scales ($t \approx \mu \text{s}$) may transport and de-confine energetic alphas before their energy is transferred to the thermal plasma, leading to inefficient plasma self-heating and localized damage to the plasma chamber wall. Accurate prediction of alpha-particle transport and losses is therefore crucial in integrated modeling workflows used for fusion reactor design and plasma scenario optimization.
\par 
In the absence of operating burning-plasma experiments producing net energy gain from D–T fusion reactions, predictive studies of energetic-particle transport currently rely almost entirely on numerical simulations. However, high-fidelity simulations of AE-driven energetic-particle transport remain computationally demanding as they entail solving multiple coupled nonlinear partial differential equations in 3-D geometry. For context, a single nonlinear simulation of Alfvén eigenmode instability driven energetic particle transport for ITER, using a medium-fidelity code FAR3d \cite{FAR3d_review}, requires approximately $250-300$ node hours with GPU parallelization.  The high computational cost and time required to run these nonlinear codes makes them impractical to be used in integrated modeling workflows for iterative calculations of fusion reactor design, necessitating the development of fast surrogate models. 
\par
Machine-learning surrogate models have recently emerged as efficient tools for accelerating computationally intensive plasma-physics calculations. Applications include surrogate models for fishbone instabilities \cite{FishboneML2024}, internal kink modes \cite{SGTC}, and pedestal peeling–ballooning stability \cite{KARHU2025}. However, to our knowledge, no machine learning surrogate model has yet been developed for Alfvén eigenmode (AE)-driven EP transport — despite AE activity being one of the most significant threats to EP confinement in burning plasma devices. In this paper, we present, to our knowledge, the first such surrogate model, trained on nonlinear FAR3d simulations (validated with JET experimental observations in D-T fusion plasmas \cite{Garcia2024,Varela_2025,Varela_2026}) of AE-driven EP transport fluxes for ITER. The objective of the surrogate is to replace repeated evaluations of the nonlinear energetic-particle transport response within integrated transport calculations rather than the full temporal evolution of Alfvén eigenmode growth.
\par
We selected ITER as the reference device for surrogate model development because it represents the international benchmark for reactor-scale burning plasma operation. Furthermore, the nonlinear FAR3d simulations used in this work have been documented in Ref.~\cite{Spong_2025}, providing a well-characterized reference scenario for the development and future comparison of surrogate models for energetic-particle transport. In addition to fusion-born alpha particles, ITER plasmas will contain energetic beam ions generated by neutral beam injection. The presence of multiple energetic-particle populations modifies both the drive and damping of Alfvén eigenmodes \cite{Zweben_TFTR}. Accordingly, the nonlinear FAR3d simulations used for surrogate development include both energetic alpha particles and beam ions, providing a reactor-relevant multi-species training dataset. 
\par
Beyond machine learning, several fast reduced models have been developed for EP transport prediction. The critical-gradient approach, built on the Trapped-Gyro-Landau-Fluid (TGLF) framework, was first applied to AE-driven EP diffusion by Sheng et al. \cite{Sheng2017} and later extended to ITER scenarios by Bass and Waltz \cite{Bass2020}. A separate resonance-broadened quasilinear framework, the "kick model," provides phase-space-resolved transport probability matrices and has been integrated into the TRANSP/NUBEAM code for time-dependent, predictive tokamak simulations \cite{Podesta2019, Gorelenkov2018}. Carlevaro et al. have developed one-dimensional reduced models from first-principles theory \cite{Carlevaro2022}, to capture non-diffusive, avalanche-like EP transport in ITER-relevant scenarios, benchmarked against nonlinear hybrid LIGKA/HAGIS simulations. While these reduced models offer lower computational costs relative to direct nonlinear simulation, they rely either on simplified critical-gradient/quasilinear diffusion closures (TGLF-EP, kick model, RBQ) or on a reduced-dimensionality mapping onto an equivalent 1D system (Carlevaro et al.), rather than learning the transport response directly from nonlinear simulation data. By training an ML-based surrogate directly on nonlinear simulations, the present approach retains the complex nonlinear dependencies contained in the underlying transport calculations without requiring a reduced, explicit transport closure, while enabling rapid evaluation of energetic-particle transport for integrated modeling applications.
\par
\begin{figure}[t]
\centering
\includegraphics[width=\linewidth]{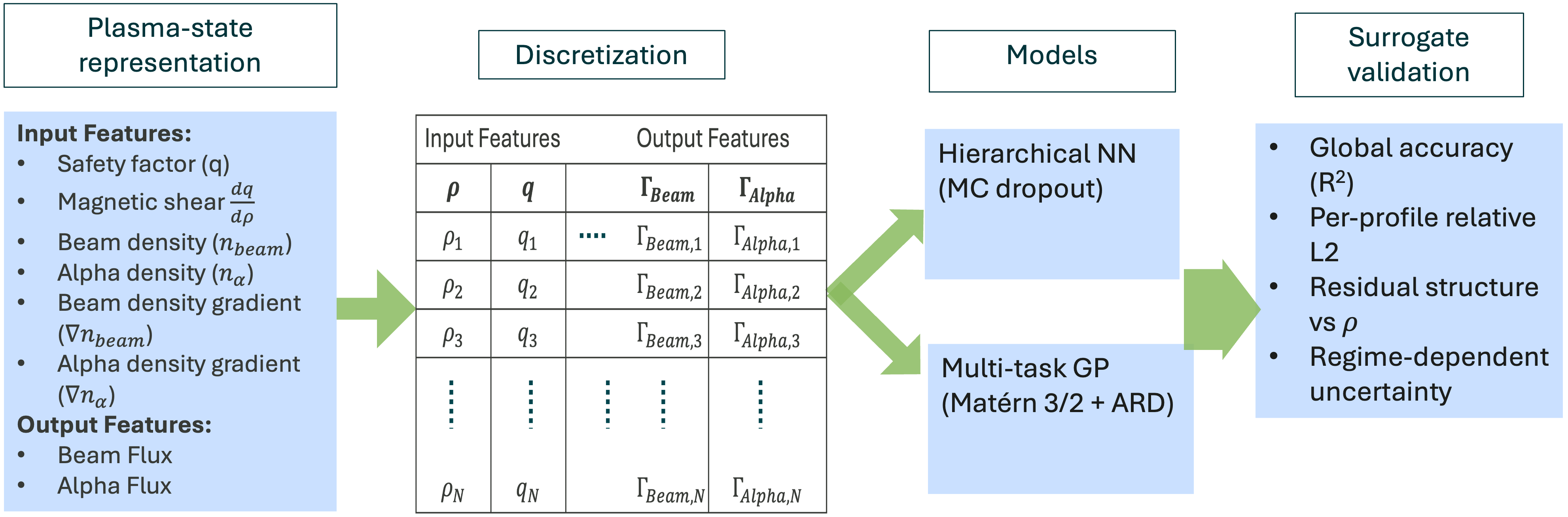}
\caption{Overview of the machine-learning surrogate development workflow.}
\label{fig:workflow}
\end{figure}

Figure~\ref{fig:workflow} provides an overview of the surrogate-model development workflow adopted in this work. Starting from the outputs from nonlinear FAR3d simulations of energetic-particle transport, the transport profiles are converted into a plasma-state representation through spatial discretization and subsequently used to train the GP and hierarchical NN surrogate models. The trained surrogate models are then assessed using complementary metrics that quantify prediction accuracy, profile reconstruction, and predictive uncertainty.
\par
\begin{figure}[h]
    \centering
    \begin{subfigure}[b]{0.48\textwidth}
        \centering
        \includegraphics[width=\textwidth]{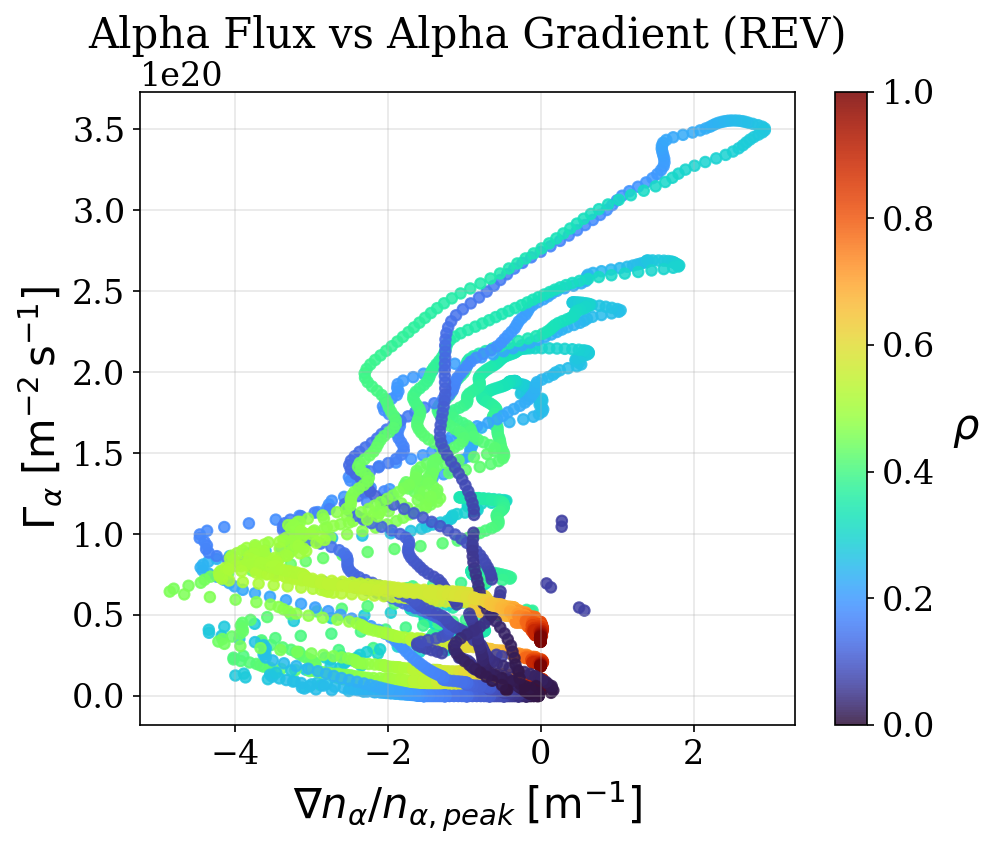}
        \caption{}
        \label{fig:fig1}
    \end{subfigure}
    \hfill
    \begin{subfigure}[b]{0.48\textwidth}
        \centering
        \includegraphics[width=\textwidth]{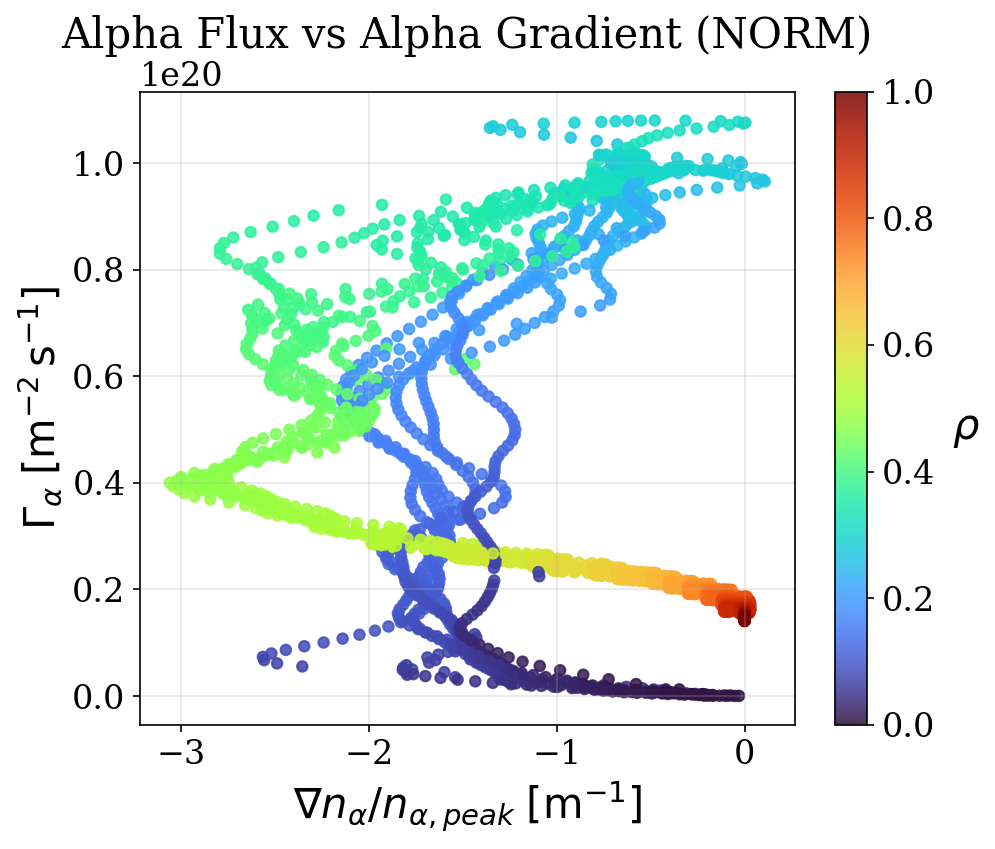}
        \caption{}
        \label{fig:fig2}
    \end{subfigure}
    \caption{Variation of alpha particle transport flux ($\Gamma_\alpha$) with normalized alpha particle density gradient ($\nabla n_{\alpha}/n_{\alpha,peak}$) for ITER steady state scenarios with (a) reverse-shear (REV) and (b)  monotonic (NORM) q-profiles. Data points are colored by the radial coordinate $\rho$ (square-root of normalized toroidal flux) at which each transport flux value is sampled.}
    \label{fig:both}
\end{figure}

In this work, we develop Gaussian process (GP) and neural-network (NN) surrogate models for predicting energetic beam and alpha-particle transport fluxes. The models are trained with data from nonlinear FAR3d simulations of an ITER steady-state scenario. The energetic-particle transport fluxes in ITER exhibit a strongly nonlinear dependence on the corresponding energetic-particle density gradients, as illustrated by the representative flux--gradient relationships shown in Fig.~\ref{fig:both}. This nonlinear behavior varies across the plasma radius and between the reversed- and monotonic-shear operating scenarios, motivating the use of machine-learning surrogate models capable of learning the transport response directly from the nonlinear FAR3d simulations. Gaussian processes (GPs) and neural networks (NNs) represent two complementary classes of nonlinear surrogate models capable of learning complex relationships between the plasma state and energetic-particle transport. GPs employ a probabilistic kernel-based formulation that naturally provides Bayesian predictive uncertainty and interpretable feature relevance through Automatic Relevance Determination (ARD). In contrast, NNs learn complex nonlinear representations directly through hierarchical nonlinear transformations of the input features. The two surrogate approaches are compared to assess their relative strengths for learning the nonlinear mapping between the local plasma state and the corresponding energetic-particle transport fluxes. Particular emphasis is placed on predictive accuracy, transport-profile reconstruction, and uncertainty quantification across different transport regimes, providing a systematic assessment of their suitability for reduced energetic-particle transport modeling. Quantifying predictive uncertainty is particularly important for integrated modeling applications, where surrogate predictions may be applied in plasma regimes that are only sparsely represented in the available training data, thereby providing a measure of confidence in the predicted transport response.
\par


\par
The remainder of this paper is organized as follows. Section~2 describes the generation of the nonlinear FAR3d dataset and the construction of the surrogate training dataset. Section~3 presents the flux-variability analysis used to validate the plasma-state representation adopted for surrogate modeling. Section~4 describes the hierarchical neural-network and multitask Gaussian process surrogate architectures. Section~5 evaluates the global prediction accuracy and computational performance of the surrogate models (in 5.1 and 5.2), while Sections~5.3 and~5.4 examine their profile reconstruction capability and predictive uncertainty across different nonlinear transport regimes. Finally, Section~6 summarizes the main conclusions of this work and outlines directions for future research.


\section{Data generation}\label{section:radial_discretization}
The nonlinear FAR3d simulations provide the time-dependent evolution of energetic-particle-driven Alfvén eigenmodes (AEs) and the corresponding energetic particle transport fluxes used to construct the training dataset for an ITER steady-state operating scenario. FAR3d simulations were performed on two magnetic equilibrium configurations for an ITER steady-state operation scenario: one with a reversed-shear $q$-profile, another with a monotonic $q$-profile. The complete nonlinear simulation dataset comprising both simulations is reported in Ref.~\cite{Spong_2025}. The paper reports significant AE activity and associated alpha particle transport computed in ITER steady-state scenario with reversed-shear q-profile, which reduces by $50\%$ for the ITER scenario with monotonic q-profile. Since the objective of the present work is to construct a surrogate model for energetic-particle transport rather than the temporal evolution of the instability itself, the selection of the training data requires careful consideration. While the following approach is demonstrated here for building a surrogate using FAR3d simulations of ITER, the method is generally applicable for surrogate model development in other devices and using data generated from other simulation models.

\par
\paragraph{Output data from nonlinearly saturated state:}Although the linear growth phase of Alfvén eigenmodes is physically important for establishing the instability, it produces relatively little energetic-particle transport because the mode amplitudes remain small and the energetic-particle profiles undergo only negligible redistribution. Moreover, the transport response of greatest relevance to integrated modeling corresponds to the nonlinear saturated regime, where nonlinear wave–particle interactions lead to appreciable energetic-particle redistribution and transport. Accordingly, for surrogate-model development, only data from the nonlinear saturated phase of the simulations were retained. In the cases considered here, this phase corresponds to a statistically stationary transport regime (often referred to as the soft-MHD regime), in which the energetic-particle fluxes fluctuate around a quasi-steady state without strong bursting activity or mode chirping. This regime is therefore well suited for constructing a steady-state transport surrogate. The present surrogate is consequently intended for the range of EP-driven mode amplitudes and nonlinear behavior represented by these simulations and does not presently encompass strongly driven regimes characterized by bursting or large-scale relaxation events. This choice is also consistent with existing reduced energetic-particle transport models, which provide transport closures for integrated modeling rather than resolving the complete temporal evolution of Alfvén eigenmode growth and nonlinear relaxation.
\par
\paragraph{Training vs testing data split:} Each simulation was performed on four compute nodes (16 GPUs). During the nonlinear saturated phase, retained transport profiles were written approximately every 14,000 simulation time steps, corresponding to approximately 10.5 hours of wall-clock simulation time between successive output dumps. For the reversed-shear ITER simulation, a total of 15 output dumps were generated, four of which were required solely to evolve the plasma from the linear growth phase into the nonlinear saturated state before transport profiles suitable for surrogate training could be extracted. The remaining 11 output dumps were retained for surrogate-model development. For the normal-shear simulations, we retained 9 output dumps in the nonlinear saturated state from the 15 dump files that were written out. From these retained output dumps, a total of 20 transport profiles were used for surrogate development. Sixteen profiles were assigned to the training dataset, while four profiles (two from the reversed-shear case and two from the monotonic-shear case) were reserved exclusively for testing. This profile-wise partitioning ensured that the surrogate models were evaluated on complete transport profiles that were not used during training.
\paragraph{Transport-flux normalization:}
The beam-ion and alpha-particle transport fluxes were scaled using species-dependent characteristic particle-flux scales constructed from the peak energetic-particle density and characteristic velocity,
\begin{equation}
    \Gamma_{s,0}^{*}
    =
    n_{s,\mathrm{peak}} v_{\mathrm{th},s},
    \qquad
    v_{\mathrm{th},s}
    =
    \sqrt{\frac{k_B T_{s,\mathrm{peak}}}{m_s}},
\end{equation}
where $s$ denotes the beam-ion or alpha-particle species. An additional factor of $10^{-5}$ was applied to these characteristic flux scales to maintain numerically well-conditioned target magnitudes during surrogate-model 
training. The effective reference fluxes used for normalization are therefore
\begin{equation}
    \Gamma_{\mathrm{beam},0}
    =
    1.1714\times10^{20}\ \mathrm{m^{-2}\,s^{-1}},
    \qquad
    \Gamma_{\alpha,0}
    =
    5.5857\times10^{19}\ \mathrm{m^{-2}\,s^{-1}}.
\end{equation}
The transport fluxes supplied to the surrogate models are consequently defined as $\Gamma_{s,\mathrm{scaled}} =
    \Gamma_s/\Gamma_{s,0}$.
This constant rescaling changes only the numerical magnitude of the regression targets and does not alter their relative spatial or temporal variation. Unless stated otherwise, all transport fluxes shown in the surrogate-model figures are presented in these normalized units.
\paragraph{Simulation output labeling:} The nonlinear saturated phase is further divided into early- and late-saturation regimes, corresponding approximately to 0.3–0.45 ms and 0.5–0.6 ms, respectively, in the nonlinear mode evolution shown in Figs.~10 and 11 of Ref.~\cite{Spong_2025}. The individual simulation outputs retained from this phase are identified throughout this work by their corresponding dump indices, with indices 14--18 denoting the early-saturation regime and 19--24 the late-saturation regime. Although the transport fluxes evolve through the self-consistent evolution of the plasma profiles, the surrogate model assumes that the instantaneous transport fluxes can be parameterized by the current plasma state. The validity of this assumption is examined through the flux-variability analysis presented in the following section.
\par
\paragraph{Radial discretization and feature-space construction:} To increase the amount of data available for surrogate training and testing, we discretized the radial profiles at every dump time and used the spatially discretized feature space containing local values of spatial coordinate ($\rho$), safety factor (q), magnetic shear ($\hat{s}$), EP densities ($n_{\mathrm{beam}}, n_{\mathrm{\alpha}}$) and EP density gradients ($\nabla n_{\mathrm{beam}}, \nabla n_{\mathrm{\alpha}}$). This discretization method created a dataset of about 9180 data points showing local values of fluxes as a function of a seven-dimensional input feature space such that the surrogate model built will capture the mapping \texttt{F}, defined by: 
\begin{equation}
\left(
\Gamma_{\mathrm{beam}},
\Gamma_{\alpha}
\right)
=
F\!\left(
\rho,\,
q,\,
\hat{s},\,
n_{\mathrm{beam}},
n_{\alpha},\,
\nabla n_{\mathrm{beam}},\,
\nabla n_{\alpha}
\right).
\label{eq:surrogate_mapping}
\end{equation}

The mapping described in Eq. \ref{eq:surrogate_mapping} will be considered unique if similar plasma states in feature space consistently produce similar transport fluxes — a necessary condition for surrogate reliability. We will assess this uniqueness of surrogate mapping in the following section. 

\section{Validation of input feature representation}\label{section:flux-consistency}
\begin{figure*}[htb!]
    \centering
    \begin{subfigure}[t]{0.45\textwidth}
        \centering
        \includegraphics[width=1.05\textwidth]{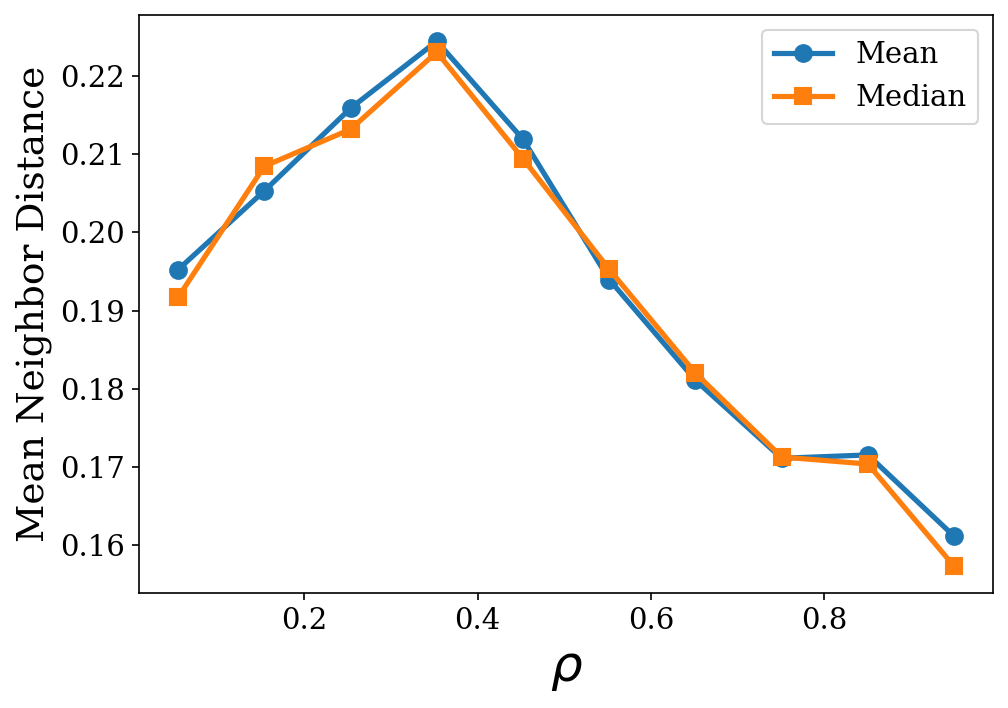}
        \caption{}
        \label{fig:variability_rho}
    \end{subfigure}
    \hfill
    \begin{subfigure}[t]{0.45\textwidth}
        \centering
        \includegraphics[width=1.05\textwidth]{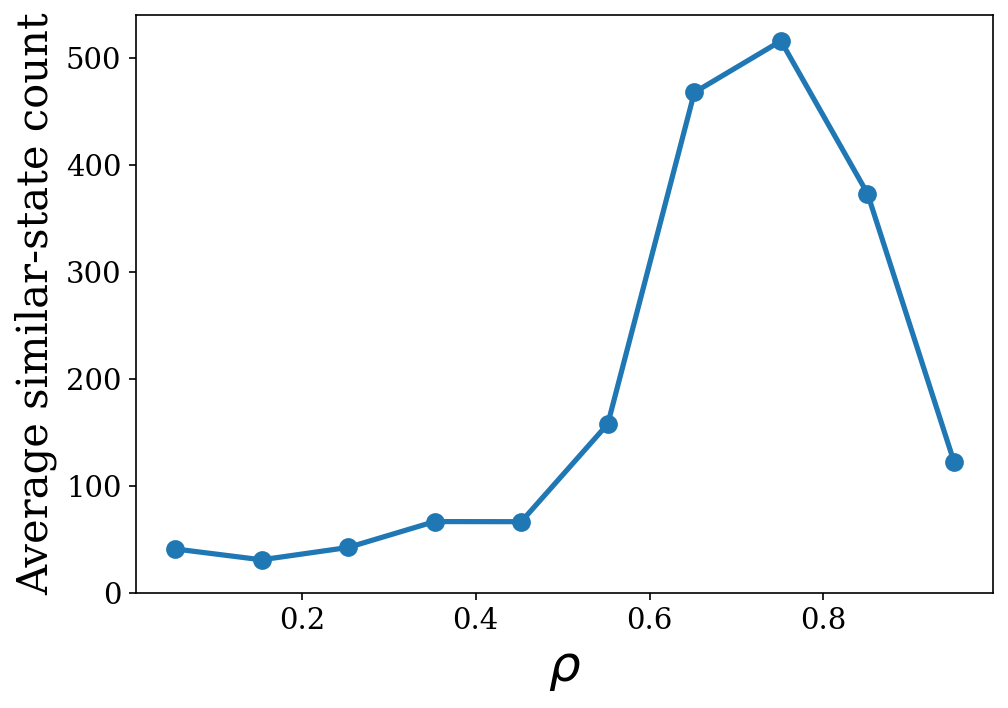}
        \caption{}
        \label{fig:similar_state_count}
    \end{subfigure}

    \vspace{0.5em}

    \begin{subfigure}[t]{0.45\textwidth}
        \centering
        \includegraphics[width=1.05\textwidth]{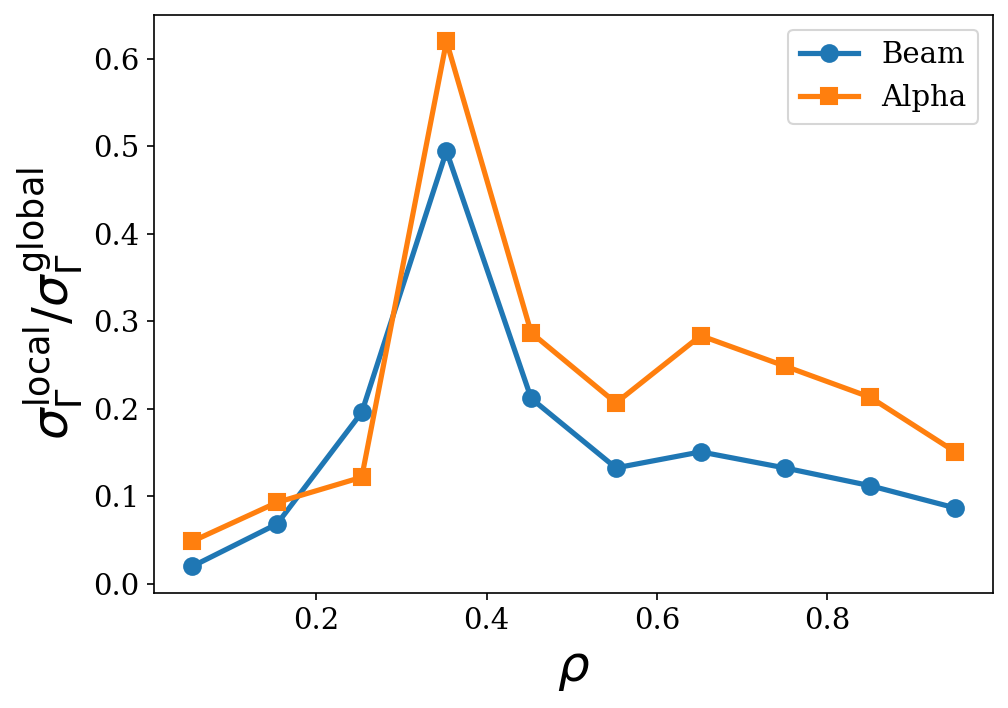}
        \caption{}
        \label{fig:neighbor_distance}
    \end{subfigure}
    \hfill
    \begin{subfigure}[t]{0.45\textwidth}
        \centering
        \includegraphics[width=1.05\textwidth]{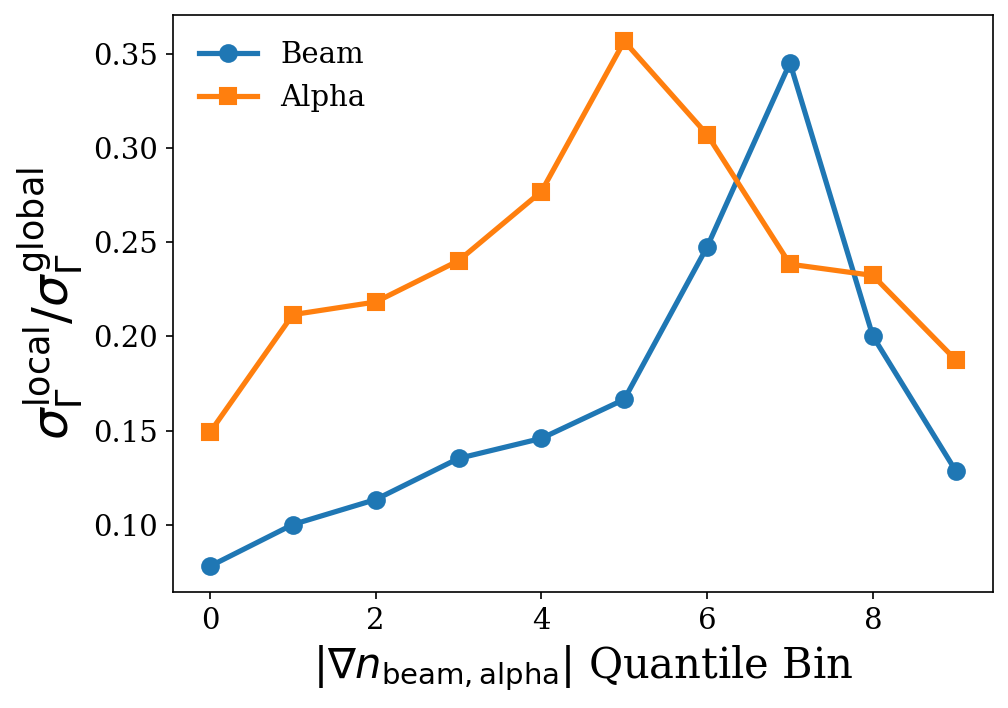}
        \caption{}
        \label{fig:variability_gradient}
    \end{subfigure}

    \caption{Assessment of the surrogate state representation using nearest-neighbor 
    analysis in the normalized seven-dimensional feature space (Eq.~\ref{eq:surrogate_mapping}). 
    (a) Mean and median nearest-neighbor distance in the normalized feature space as a 
    function of $\rho$, validating that the fixed-radius threshold of 0.3 consistently 
    exceeds the typical inter-point spacing across the radial domain. 
    (b) Average number of similar cross-profile states per radial location identified 
    within the feature-space neighborhood (Euclidean distance $< 0.3$). 
    (c) Ratio of local-to-global flux variability ($\sigma_{\Gamma}^{\mathrm{local}} / 
    \sigma_{\Gamma}^{\mathrm{global}}$) as a function of normalized radius ($\rho$) for 
    beam and alpha particle transport.
    (d) Local-to-global flux variability as a function of EP density gradient strength 
    quantile bin ($|\nabla n_{\mathrm{beam, alpha}}|$).}
    \label{fig:surrogate_justification}
\end{figure*}

 To assess the uniqueness of the surrogate mapping described in Eq.~\ref{eq:surrogate_mapping}, we performed a local transport-flux variability analysis in the normalized seven-dimensional input feature space. By comparing EP transport fluxes at similar plasma states, this analysis evaluates whether the selected input features provide a sufficiently informative representation of the underlying transport dynamics.
 \paragraph{Identification of neighboring states:} For each plasma state, we identified neighboring states within a fixed-radius neighborhood (Euclidean distance $< 0.3$), excluding neighbors originating from the same simulation profile to avoid trivial self-similarity. The threshold of $0.3$ was selected based on the mean nearest-neighbor distance in the normalized feature space, which ranges from 0.16 to 0.23 across the radial domain (Fig. \ref{fig:surrogate_justification} (a)). This ensures that the neighborhood radius exceeds the typical inter-point spacing throughout the dataset while remaining sufficiently local to capture only genuinely similar plasma states. 
 \paragraph{Quantification of flux variability within the identified neighboring states:} We computed the ratio of the standard deviation of the transport fluxes corresponding to the identified neighboring plasma states in feature-space, with respect to the standard deviation of the transport fluxes over the complete dataset. 
 The lower values of this ratio indicate a more unique mapping between the plasma state and the corresponding transport flux, whereas larger values imply increased variability among similar plasma states.
\paragraph{Flux variability across the radial domain and gradient regimes:} The results indicate that the local-to-global flux variability ratio remains below 0.3 across most of the radial domain for both beam and alpha fluxes as shown in Figure \ref{fig:surrogate_justification} (c), confirming that the surrogate mapping (Eq. \ref{eq:surrogate_mapping}) is reasonably unique over most of the sampled manifold. The largest deviations in the local-to-global flux variability ratio are observed near the region of peak energetic-particle transport ($\rho \sim 0.35$). As shown in Fig. \ref{fig:surrogate_justification} (b), this region is also characterized by a lower density of similar states in feature space compared to the outer radial region. This suggests that reduced feature-space density contributes to the larger variability observed near the transport maximum. Figure \ref{fig:surrogate_justification} (d) further shows that the largest local-to-global flux variability occurs at intermediate-to-high EP gradient strengths for both beam and alpha particle fluxes. It is also shown in both Figs. \ref{fig:surrogate_justification} (c) and (d) that the overall variability among similar feature states is lower for beam fluxes than for the alpha particle fluxes. This suggests that the selected feature set provides a more unique parameterization of beam transport and may therefore lead to a more accurate surrogate model for beam flux predictions, and more accurate predictions of transport fluxes where the gradients are small.
 \par
Overall, the local transport-flux variability analysis shows that the states exhibiting high values of local-to-global flux variability ratios ($>0.3$), corresponding to local flux variability greater than $30\%$ of the global flux variability, constitute only approximately $10\%$ of the dataset for the beam transport fluxes and $17\%$ of the dataset for the alpha-particle fluxes. Moreover, these states are concentrated near the strongest transport and gradient regimes, whereas the majority of the sampled feature space exhibits substantially lower variability. This suggests that the reduced plasma-state representation provides a unique description of the transport response throughout most of the sampled manifold, while predictions in the strongest transport and gradient regimes may be more sensitive to effects not fully captured by the selected feature set. This is further examined quantitatively in section \ref{section:Discussion}.

\section{Surrogate model architectures}
As per the discretization of radial flux profiles explained in section \ref{section:radial_discretization}, the resulting nonlinear simulation dataset contains approximately 9180 local plasma states, making it sufficiently large to support neural-network training while remaining computationally feasible for exact Gaussian process regression. This provides an opportunity to compare two complementary classes of nonlinear surrogate models: a hierarchical neural-network surrogate and a multitask Gaussian process surrogate. 

Although the two surrogate models employ different strategies for incorporating global information beyond the local plasma state, both were optimized to achieve comparable predictive accuracy, thereby enabling differences in predictive behavior and uncertainty quantification to be assessed independently of overall model performance. The input representations adopted by the two surrogate models are summarized in Table~\ref{tab:input_features}, while their individual architectures are described below.
\subsection{NN surrogate model}
\par
To capture the distinct transport characteristics of the inner and outer plasma regions, a hierarchical neural-network surrogate was developed consisting of two feed-forward neural networks trained sequentially for the inner (low $\rho$) and outer (high $\rho$) radial regions, with the boundary chosen at the radial co-ordinate $\rho=0.5$. Preliminary investigations showed that training a single neural network over the entire radial domain resulted in the high-$\rho$ transport fluxes being systematically averaged because of their substantially smaller variability compared to the low-$\rho$ transport region. To overcome this limitation, the low-$\rho$ network was first trained using the seven local plasma-state variables $(\rho,q,\hat{s},n_{\mathrm{beam}},n_{\alpha},\nabla n_{\mathrm{beam}},\nabla n_{\alpha})$. The high-$\rho$ network was subsequently trained using the same seven local plasma-state variables together with the mean beam and alpha-particle transport fluxes predicted by the low-$\rho$ network. These additional inputs provide information about the overall transport level established in the inner plasma and allow the high-$\rho$ network to account for the dependence of outer-region transport on the overall transport level established in the low-$\rho$ region.
\par
Both the low- and high-$\rho$ networks employed the same fully connected feed-forward architecture consisting of an input layer, two hidden layers with 64 neurons each, and an output layer containing two neurons corresponding to the beam and alpha-particle transport fluxes. Rectified Linear Unit (ReLU) activation functions were used for the hidden layers, while a Softplus activation function was applied to the output layer to ensure positive transport flux predictions. A dropout layer was included after each hidden layer to randomly deactivate approximately $10\%$ of the neurons during each training iteration. This reduces overfitting by decreasing the dependence of the network on individual neurons, thereby improving the robustness of the learned transport mapping. During inference, the same dropout layers were retained to enable predictive uncertainty estimation using the Monte Carlo dropout approach.
\par
Prior to training, each input feature was standardized using the mean and standard deviation of the training dataset. The network parameters were optimized using the Adam optimizer with a learning rate of $5\times10^{-4}$. Both networks were trained for 3000 epochs using a weighted mean-squared-error loss function in which larger transport fluxes were assigned proportionally larger weights in the loss function, thereby improving the prediction of peak transport events. Gradient clipping was used during optimization to improve training stability.

\subsection{GP surrogate model}
\par
The GP surrogate was implemented using a multitask Gaussian process regression model to simultaneously predict the beam and alpha-particle transport fluxes. In contrast to the hierarchical NN surrogate, the GP employed a single regression model over the entire radial domain using the local plasma-state representation introduced in Section~\ref{section:flux-consistency} ($\rho,q,\hat{s},n_{\mathrm{beam}},n_{\alpha},\nabla n_{\mathrm{beam}},\nabla n_{\alpha}$) together with the profile-averaged beam and alpha-particle densities $(\langle n_{\mathrm{beam}}\rangle,\langle n_{\alpha}\rangle)$. The inclusion of the profile-averaged densities provides information about the overall energetic-particle inventory and instability-drive strength for each simulation dump.
\par
A Mat\'ern 3/2 covariance kernel with Automatic Relevance Determination (ARD) was employed to model the nonlinear dependence of the transport fluxes on the plasma-state variables. This kernel was selected because it provides sufficient flexibility to represent nonlinear transport behavior while avoiding the overly smooth function assumptions associated with the squared-exponential (RBF) kernel. The ARD formulation assigns an independent characteristic length scale to each input feature, thereby allowing the GP to determine the relative influence of each plasma-state parameter on the predicted transport fluxes. Correlations between the beam and alpha-particle transport fluxes were incorporated through a multitask covariance function, enabling the two transport channels to be learned simultaneously.
\par
Prior to training, all input features were standardized using the mean and standard deviation of the training dataset. The GP hyper-parameters, including the kernel length scales, output covariance, and observation noise, were optimized by maximizing the exact marginal log-likelihood using the Adam optimizer. Unlike the NN surrogate, the GP provides predictive uncertainties as part of the regression model itself through the posterior predictive variance.

\begin{table}[t]
\centering
\caption{Input representations used by the hierarchical NN and multitask GP surrogate models.}
\begin{tabular}{p{3cm}p{10.5cm}}
\toprule
\textbf{Network} & \textbf{Input representation} \\
\midrule

\multirow{2}{*}{Hierarchical NN}
&
{\raggedright
\textbf{Low-$\rho$:}
Seven local plasma-state variables
$(\rho,q,\hat{s},
n_{\mathrm{beam}},
n_{\alpha},
\nabla n_{\mathrm{beam}},
\nabla n_{\alpha})$.\par}
\\[0.4em]

&
{\raggedright
\textbf{High-$\rho$:}
Seven local plasma-state variables together with the mean predicted
beam and alpha-particle transport fluxes obtained from the
low-$\rho$ network.\par}
\\

\midrule

Multitask GP
&
{\raggedright
\textbf{Entire domain:}
Seven local plasma-state variables together with the
profile-averaged beam and alpha-particle densities,
$\langle n_{\mathrm{beam}}\rangle$ and
$\langle n_{\alpha}\rangle$.\par}
\\

\bottomrule
\end{tabular}
\label{tab:input_features}
\end{table}
\section{Prediction Accuracy and Uncertainty Quantification}\label{section:Discussion}
The performance of the GP and NN surrogate models is evaluated using complementary metrics that assess global prediction accuracy, profile reconstruction quality, computational efficiency, and predictive uncertainty across different nonlinear transport regimes. It is important to note that both surrogate models were optimized to provide comparable predictive accuracy, allowing the subsequent comparisons to focus primarily on differences in predictive behavior and uncertainty quantification.
\subsection{Global Prediction Accuracy}
\begin{figure}[htb!]
    \centering
    
    \begin{subfigure}{0.48\textwidth}
        \centering
        \includegraphics[width=\textwidth,height=6cm]{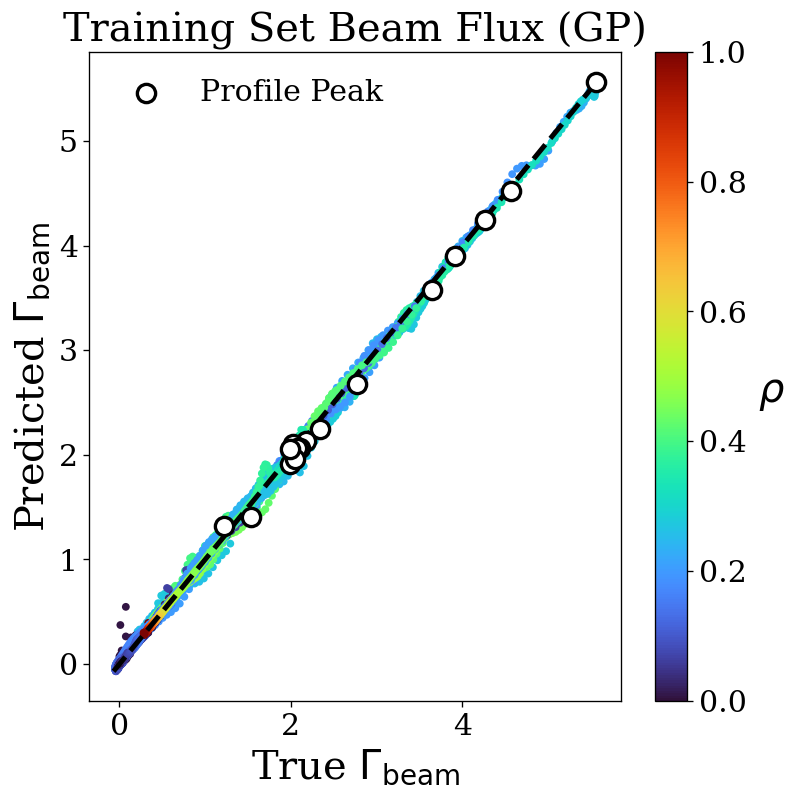}
        \caption{}
    \end{subfigure}
    \hfill
    \begin{subfigure}{0.48\textwidth}
        \centering
        \includegraphics[width=\textwidth,height=6cm]{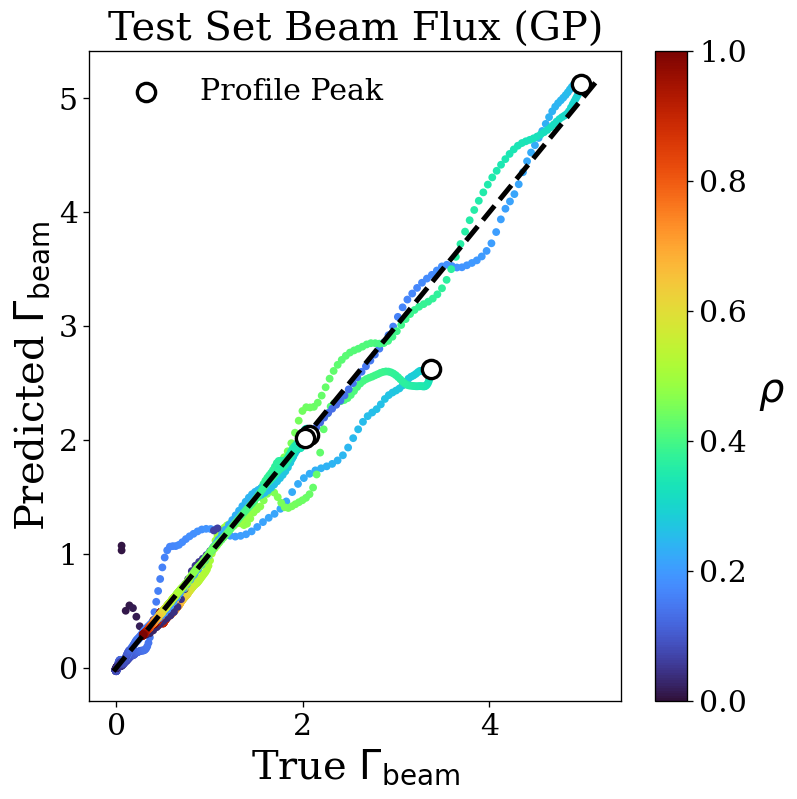}
        \caption{}
    \end{subfigure}

    \vspace{0.4cm}

    \begin{subfigure}{0.48\textwidth}
        \centering
        \includegraphics[width=\textwidth,height=6cm]{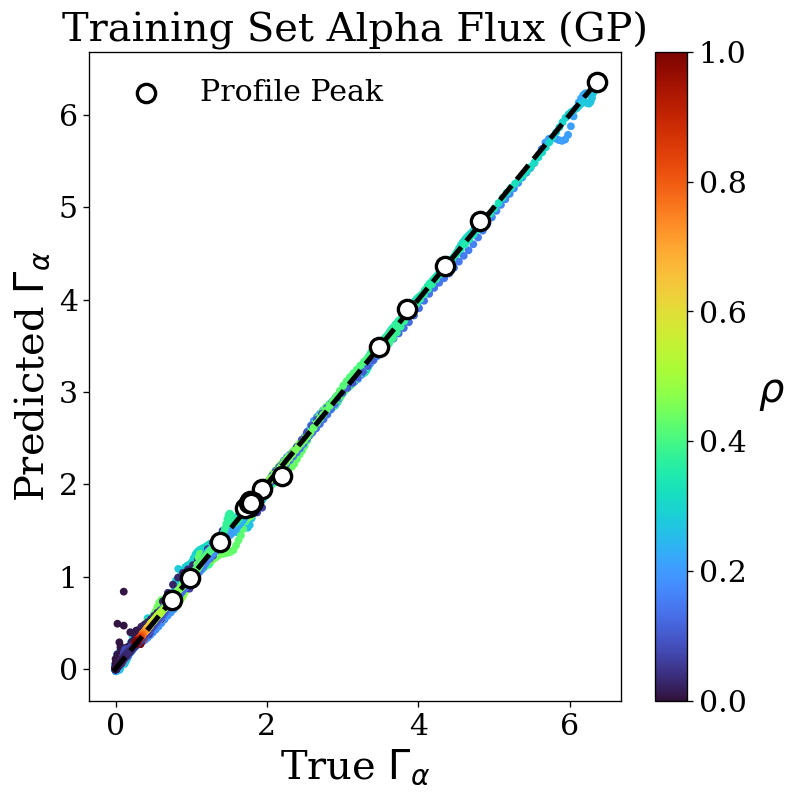}
        \caption{}
    \end{subfigure}
    \hfill
    \begin{subfigure}{0.48\textwidth}
        \centering
        \includegraphics[width=\textwidth,height=6cm]{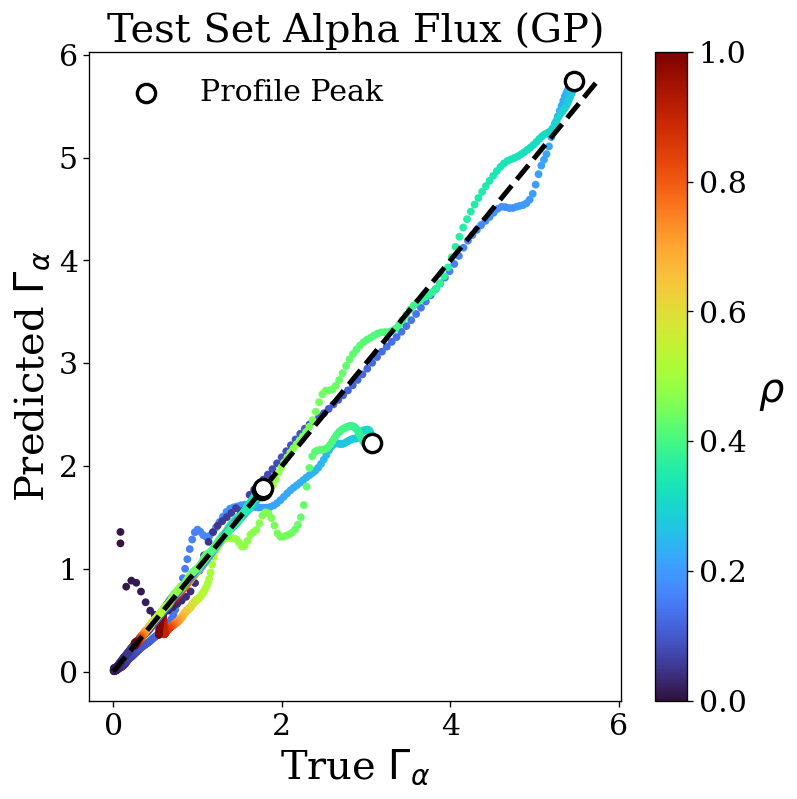}
        \caption{}
    \end{subfigure}
    \caption{Scatter plots showing energetic alphas and beam particle fluxes predicted by the multitask-GP surrogate model for training data (subfigures (a) and (c)) and for test data (subfigures (b) and (d)). The colormap indicates the radial location of these fluxes and the black circles represent the peak flux values for each radial flux profile. The dashed diagonal line indicates a perfect match of predicted fluxes to actual values. The transport fluxes are shown in normalized units as defined in Sec.~\ref{section:radial_discretization}. }
    \label{scatterplot:GP}
\end{figure}

\begin{figure}[htb!]
    \centering
    
    \begin{subfigure}{0.48\textwidth}
        \centering
        \includegraphics[width=\textwidth,height=6cm]{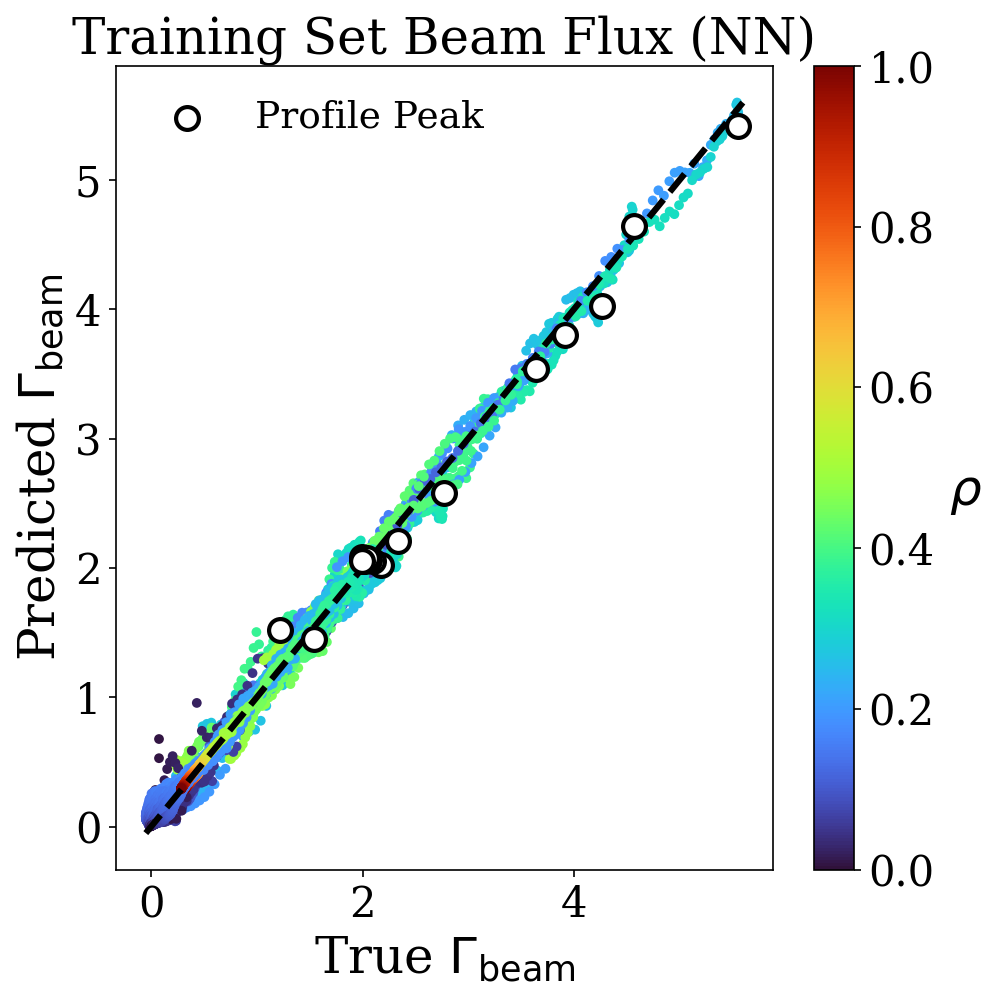}
        \caption{}
    \end{subfigure}
    \hfill
    \begin{subfigure}{0.48\textwidth}
        \centering
        \includegraphics[width=\textwidth,height=6cm]{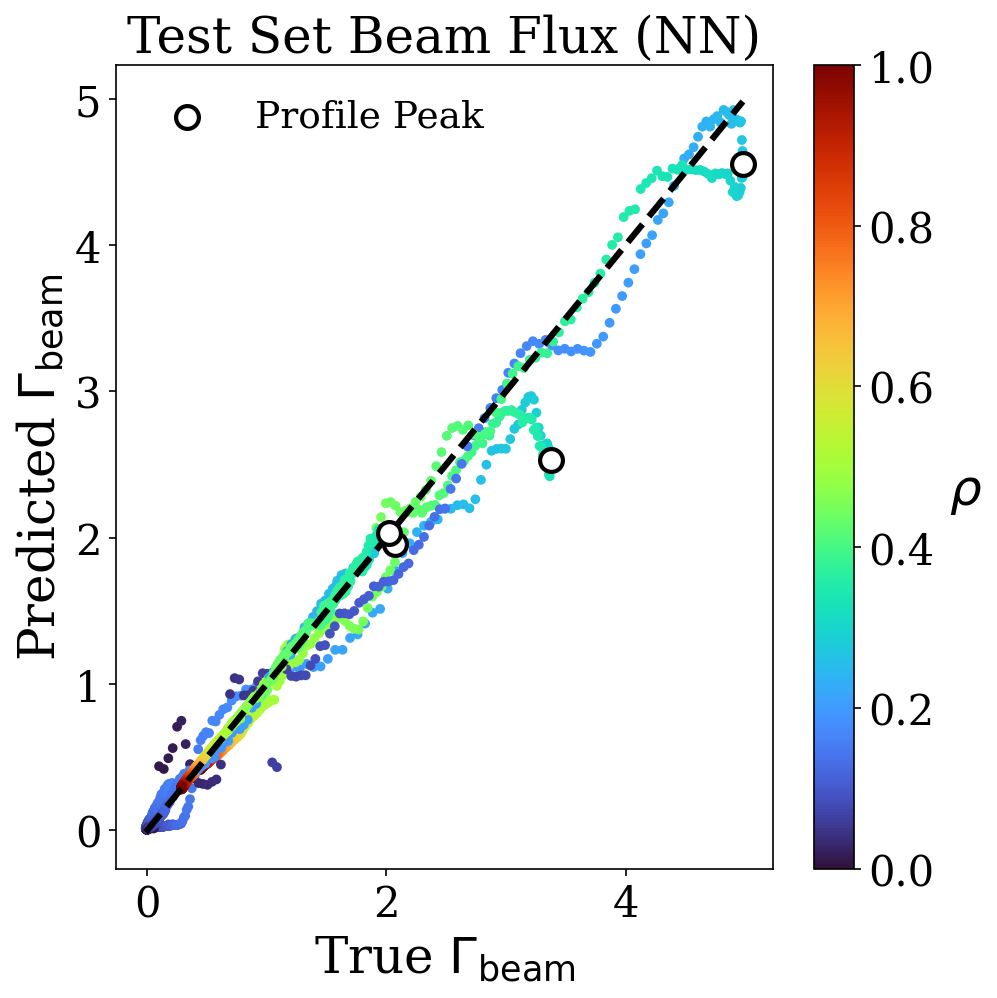}
        \caption{}
    \end{subfigure}

    \vspace{0.4cm}

    \begin{subfigure}{0.48\textwidth}
        \centering
        \includegraphics[width=\textwidth,height=6cm]{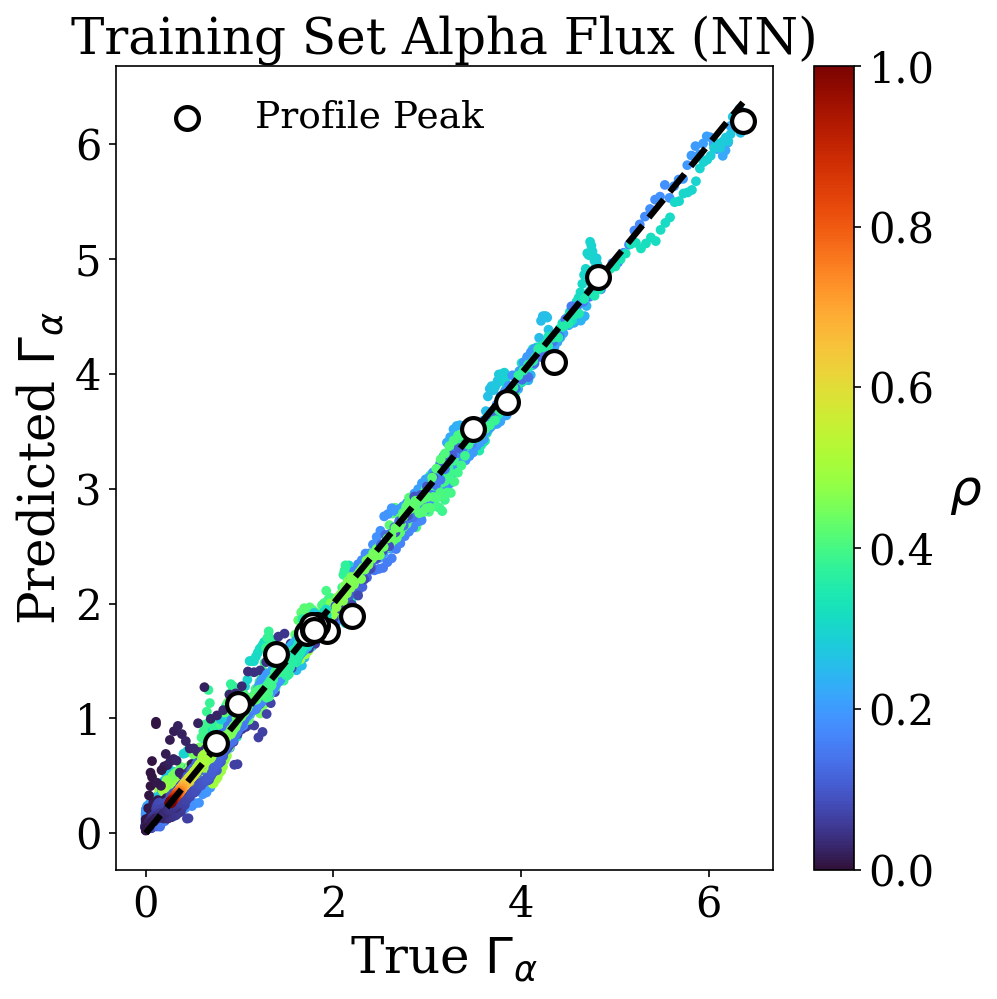}
        \caption{}
    \end{subfigure}
    \hfill
    \begin{subfigure}{0.48\textwidth}
        \centering
        \includegraphics[width=\textwidth,height=6cm]{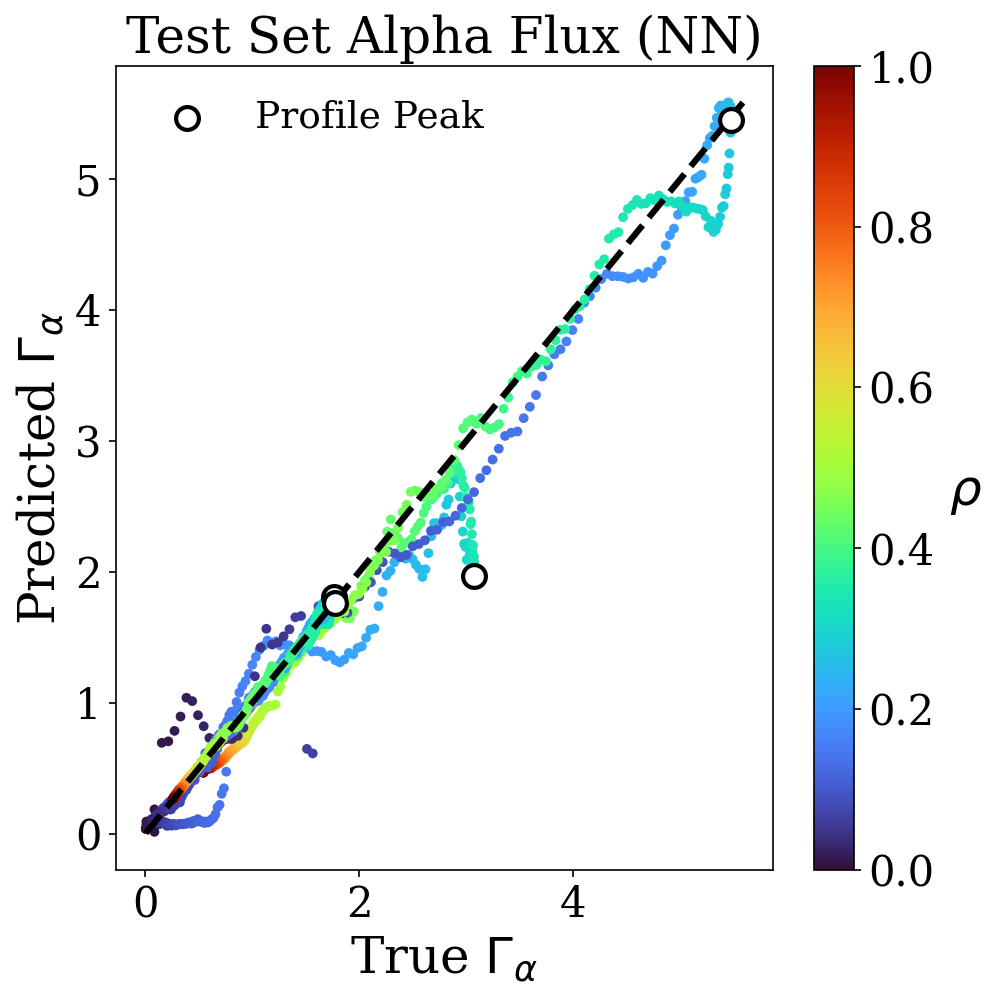}
        \caption{}
    \end{subfigure}
    \caption{Scatter plots showing energetic alphas and beam particle fluxes predicted by the neural network based surrogate model for training data (subfigures (a) and (c)) and for test data (subfigures (b) and (d)). The colormap indicates the radial location of these fluxes and the black circles represent the peak flux values for each radial flux profile. The transport fluxes are shown in normalized units as defined in Sec.~\ref{section:radial_discretization}.}
    \label{scatterplot:NN}
\end{figure}

Figures \ref{scatterplot:GP} and \ref{scatterplot:NN} show the plots of predicted versus actual values of the beam and alpha particle fluxes calculated, respectively, by the multi-task GP and the NN based surrogate models, for both testing and training datasets. These figures show that both GP and NN capture the global trend extremely well while also preserving the radial structure and peak locations of the transport flux profiles. On careful observation, the GP model predictions follow the diagonal line somewhat more closely, indicating slightly improved point-wise agreement with the target fluxes for both the training and testing datasets. It is also important to note that, despite the larger flux variability observed near the peak transport regions associated with strong EP density gradients, as identified by the local transport-flux variability analysis, both models capture the peak values of the transport profiles reliably throughout most of the dataset, showing visible deviations from the diagonal for only a small number of profiles. This result is particularly encouraging, as it demonstrates that the surrogate models remain capable of reproducing the strongest transport events even in regions where similar plasma states exhibit an increased variability in transport fluxes.

\par
Table~\ref{tab:timing} compares the computational cost of generating the nonlinear transport data with the evaluation time of the trained GP and NN surrogate models. The runtime comparison is presented for the reversed-shear ITER simulation, although similar computational costs are obtained for the monotonic-shear case since the simulation setup and surrogate architectures are identical. Generating the complete set of retained transport profiles required approximately 158 wall-clock hours (2528 GPU-hours) of nonlinear simulation time. In contrast, both surrogate models reproduce the corresponding transport outputs, including predictive uncertainty estimates, in only a few seconds. This represents a reduction in computational cost of approximately five to six orders of magnitude relative to direct nonlinear FAR3d simulations, making the surrogate models well suited for integrated modeling workflows requiring repeated transport evaluations.
\begin{table}[htb!]
\centering
\caption{Comparison of the computational cost of direct nonlinear FAR3d simulations and the trained surrogate models for evaluating energetic-particle transport. The reported GP and NN run-times correspond to the prediction of all retained transport profiles from a complete ITER simulation together with their associated predictive uncertainty estimates.}
\label{tab:timing}

\begin{tabular}{p{4.2cm}p{6.8cm}c}
\toprule
\textbf{Method} & \textbf{Quantity evaluated} & \textbf{Evaluation time} \\
\midrule

Nonlinear FAR3d simulation &
One complete ITER nonlinear simulation (11 retained transport profiles) &
158 wall-clock hours\\
(2528 GPU-hours) \\

Multitask GP &
Prediction of all 11 transport profiles
(mean + posterior uncertainty) &
2.97 s \\

Hierarchical NN &
Prediction of all 11 transport profiles
(mean + MC-dropout uncertainty) &
0.19 s \\

\bottomrule
\end{tabular}
\end{table}
\par
Having established the global predictive accuracy and computational efficiency of the surrogate models, we now examine how accurately they reconstruct the radial transport profiles.
\subsection{Profile Reconstruction Accuracy}
\par
\begin{figure}[htb!]
\centering
\includegraphics[width=0.85\linewidth]{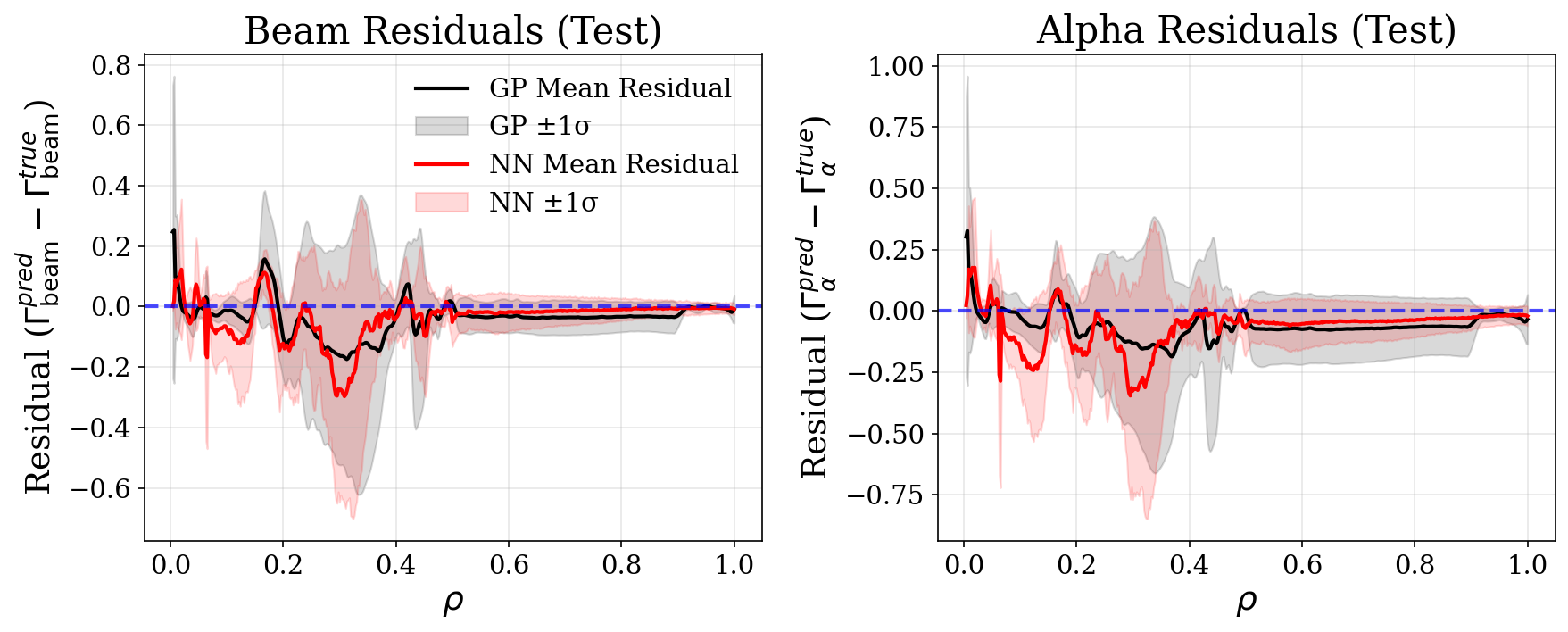}
\caption{Comparison of the residual transport flux errors, defined as $\Gamma_{\mathrm{pred}}-\Gamma_{\mathrm{true}}$, obtained from the GP and NN surrogate models for beam (top) and alpha-particle (bottom) transport fluxes. The shaded regions represent one standard deviation of the residuals calculated over all transport profiles in the training and testing datasets.}
\label{fig:residuals}
\end{figure}

 
To further compare the predictive behavior of the two surrogate models, Figure \ref{fig:residuals} shows the radial distribution of the mean residual errors for the testing dataset. As shown in Figure \ref{fig:residuals}, both models exhibit relatively small and nearly uniform residuals in the outer radial region ($\rho>0.5$), whereas larger oscillations are observed in the inner radial region ($\rho<0.5$). The largest residual amplitudes occur near the location of peak transport fluxes ($\rho\sim0.35$), which was also identified by the flux-consistency analysis as the region exhibiting the highest local flux variability. The GP model exhibits a smaller residual error near the transport peak but slightly larger errors near the plasma axis compared to the NN model. This suggests that the GP model provides a slightly more accurate reconstruction of the peak transport region, while the overall residual behavior of the two models remains comparable throughout the rest of the radial domain. Furthermore, the mean residuals remain close to zero over most of the radial domain, indicating that neither surrogate model exhibits a significant systematic bias in the predicted transport fluxes.
To identify how well these models capture the radial transport flux profile structure at each dump time, we evaluate the relative $L^2$ error for each profile using the expression:
\begin{equation}
\mathrm{Relative}\;L^2 =
\frac{
\sqrt{\displaystyle\int
\left(
\Gamma_{\mathrm{pred}}(\rho)
-
\Gamma_{\mathrm{true}}(\rho)
\right)^2
\, d\rho}
}
{
\sqrt{\displaystyle\int
\Gamma_{\mathrm{true}}^2(\rho)
\, d\rho}
}, 
\label{eq:relative_l2}
\end{equation}
where $\Gamma_{\mathrm{pred}}(\rho)$ and
$\Gamma_{\mathrm{true}}(\rho)$ denote the predicted and true
transport flux profiles, respectively, and $\rho$ is the normalized
radial coordinate. The numerator represents the $L^2$ norm of the
profile reconstruction error, while the denominator normalizes the
error by the magnitude of the true flux profile. \par

The coefficient of determination is given by

\begin{equation}
R^2
=
1
-
\frac{
\displaystyle\sum_{i=1}^{N}
\left(
\Gamma_i^{\mathrm{true}}
-
\Gamma_i^{\mathrm{pred}}
\right)^2
}
{
\displaystyle\sum_{i=1}^{N}
\left(
\Gamma_i^{\mathrm{true}}
-
\overline{\Gamma^{\mathrm{true}}}
\right)^2
}.
\label{eq:r2}
\end{equation}

where $\Gamma_i^{\mathrm{true}}$ and
$\Gamma_i^{\mathrm{pred}}$ are the true and predicted transport fluxes
at sample $i$, $\overline{\Gamma^{\mathrm{true}}}$ is the mean of the
true flux values, and $N$ is the total number of samples used in the
evaluation. An $R^2$ value of unity corresponds to a perfect prediction, whereas $R^2=0$ indicates performance equivalent to predicting the mean of the dataset. 
\par
\begin{figure}[htb!]
    \centering
    \begin{subfigure}{0.48\textwidth}
        \centering
        \includegraphics[width=\textwidth]{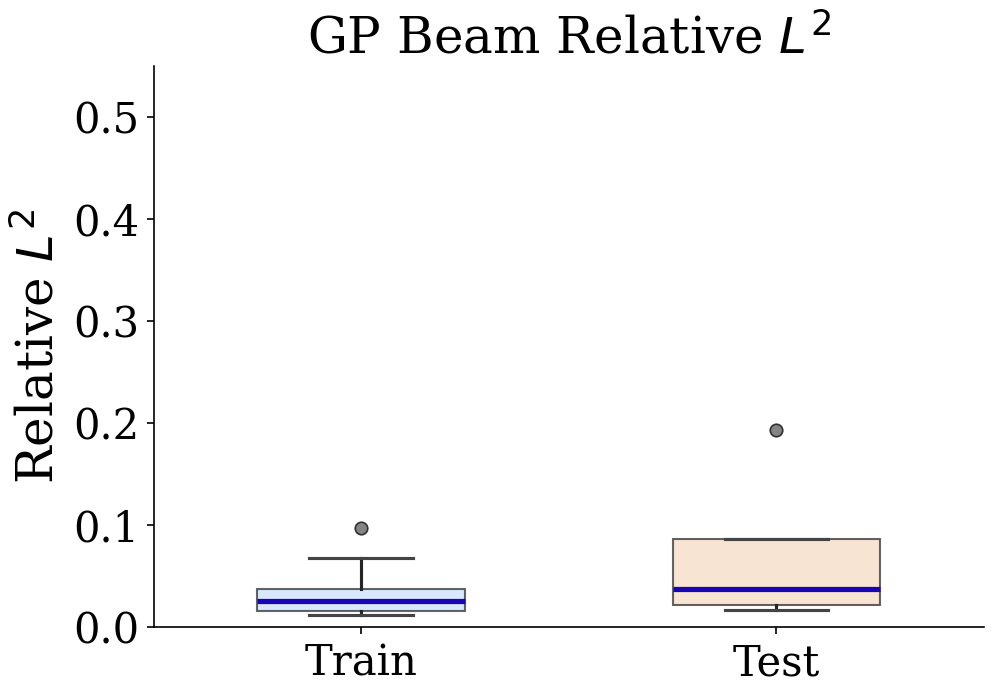}
        \subcaption{}
        \label{fig:subfig1}
    \end{subfigure}
    \hfill
    \begin{subfigure}{0.48\textwidth}
        \centering
        \includegraphics[width=\textwidth]{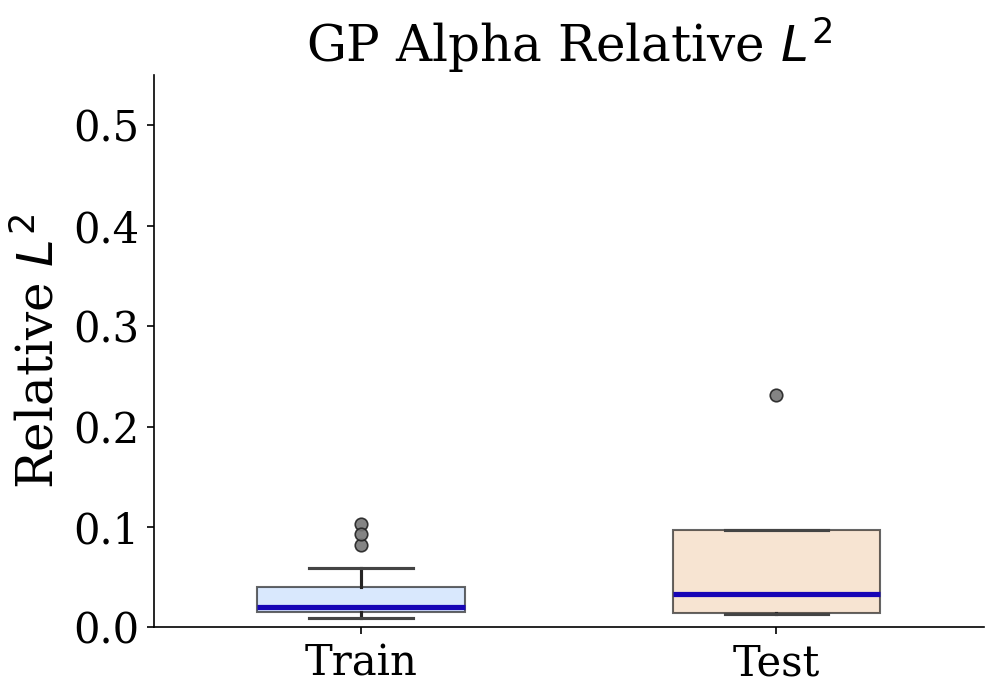}
        \subcaption{}
        \label{fig:subfig2}
    \end{subfigure}

    \vspace{0.5cm}

    \begin{subfigure}{0.48\textwidth}
        \centering
        \includegraphics[width=\textwidth]{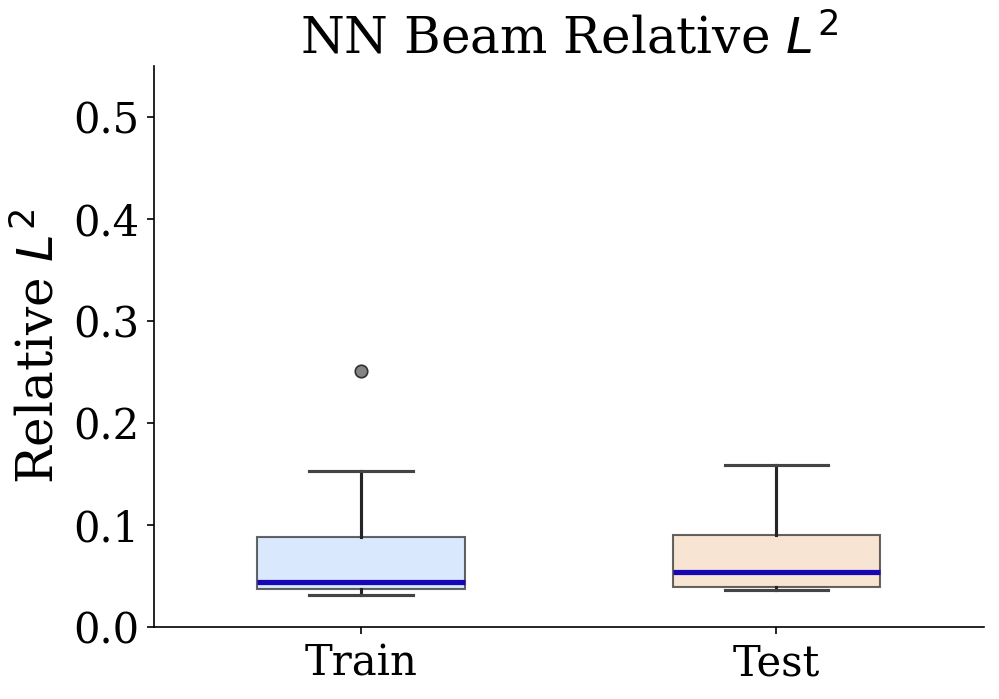}
        \subcaption{}
        \label{fig:subfig3}
    \end{subfigure}
    \hfill
    \begin{subfigure}{0.48\textwidth}
        \centering
        \includegraphics[width=\textwidth]{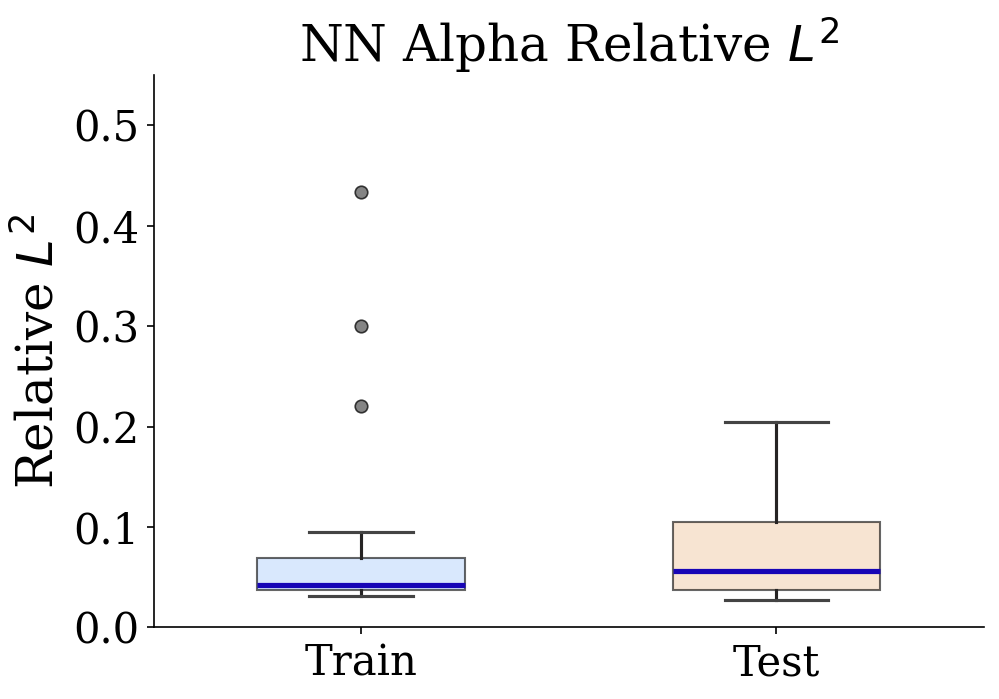}
        \subcaption{}
        \label{fig:subfig4}
    \end{subfigure}

    \caption{Boxplots of the relative $L^2$ error, computed using Eq.~\ref{eq:relative_l2}, for beam and alpha-particle flux predictions obtained from the GP (top row) and NN (bottom row) surrogate models. The colored box spans the interquartile range (IQR), corresponding to the 25th--75th percentiles of the profile-wise relative $L^2$ errors, while the horizontal line indicates the median error. The whiskers extend to the most extreme profile errors lying within the interval $[Q_1-1.5\,\mathrm{IQR},\,Q_3+1.5\,\mathrm{IQR}]$, where $\mathrm{IQR}=Q_3-Q_1$. Profiles with relative $L^2$ errors outside this interval are shown as shaded grey circles.}
    \label{fig:l2_boxplots}
\end{figure}

Figure \ref{fig:l2_boxplots} compares the distributions of profile-wise relative $L^2$ errors for beam and alpha-particle flux predictions obtained using the GP and NN surrogate models. For the training dataset, the GP model generally exhibits lower median errors and a narrower spread of profile-wise reconstruction errors than the NN model. For the testing dataset, both models exhibit comparable median relative $L^2$ errors, indicating similar profile reconstruction accuracy on unseen data. It is also important to note that for both GP and NN models, the median errors obtained on the testing dataset remain close to those observed for the training dataset indicating that neither model exhibits significant overfitting. Further, it can be observed in Fig.~\ref{fig:l2_boxplots} that a small number of outlier profiles are observed in the training dataset for both models, although the magnitude of these outliers is generally smaller for the GP model. On examining these outliers, we found that these points correspond to the density profiles calculated early in the phase of the FAR3d nonlinear simulations of ITER case with reversed shear q-profile (dump index 14,15,16). The GP model also predicts higher-error for an outlier test profile that corresponds to a dump index of 18 in the early saturation phase of the nonlinear FAR3d simulations for the same reversed shear scenario, whereas the NN model shows a wider error distribution for the testing data. 
\par
Table \ref{tab:errors} shows the quantitative comparison of the GP and NN surrogate model predictions for beam and alpha-particle transport fluxes. While both models seem to have similar prediction accuracy measured by the coefficient of determination ($R^2$), the GP model has a lower average profile prediction error ($L^2$) as compared to the NN model for both beam and alpha particle fluxes, which is also shown by the outlier profiles showing lower $L^2$ errors for the GP surrogate models compared to the NN surrogate as shown in Fig.~\ref{fig:l2_boxplots}. 

\begin{table}[htb!]
\centering
\caption{Quantitative comparison of the GP and NN surrogate models for beam and alpha-particle transport flux prediction. The relative $L^2$ error is defined by Eq.~\ref{eq:relative_l2} and the coefficient of determination $R^2$ is defined in Eq.~\ref{eq:r2}.}
\label{tab:errors}

\begin{tabular}{lcccc}
\toprule
\textbf{Metric} & \textbf{GP Train} & \textbf{GP Test} & \textbf{NN Train} & \textbf{NN Test} \\
\midrule
Beam $R^2$                  & 0.9989 & 0.9791 & 0.9952 & 0.9823 \\
Alpha $R^2$                 & 0.9991 & 0.9737 & 0.9948 & 0.9752 \\
Beam Relative $L^2$         & 0.0328 & 0.0714 & 0.0713 & 0.0768 \\
Alpha Relative $L^2$        & 0.0347 & 0.0781 & 0.0970 & 0.0865 \\
\bottomrule
\end{tabular}

\end{table}

\subsection{Regime-dependent uncertainty quantification}\label{subsection:uncertainty_quantification}
\par
To probe these outliers (difficult to predict profiles) further, we investigate how the two models capture the  uncertainties in the flux predictions for different physics regimes in the simulation, i.e., early vs later in the saturated phase of FAR3d simulations for the reversed shear ITER case. 
\paragraph{Uncertainty evaluation in GP and NN surrogates:} It is well-known that the Gaussian process model provides the prediction uncertainty as part of the model \cite{GPmethod}. For the NN surrogate, predictive uncertainties were estimated using the Monte Carlo (MC) dropout approach \cite{NNuncertainty}. During evaluation, the dropout layers were kept active and multiple forward passes were performed through the network for the same input sample. This produced an ensemble of flux predictions from which the predictive mean and standard deviation were calculated. The standard deviation was then used as a measure of uncertainty in the flux prediction. Unlike the GP uncertainty, which is obtained directly from the posterior variance of the regression process, the NN uncertainty reflects the sensitivity of the predicted fluxes to perturbations in the learned network weights.

\begin{figure}[htb!]
    \centering
    \begin{subfigure}{0.48\textwidth}
        \centering
        \includegraphics[width=\textwidth]{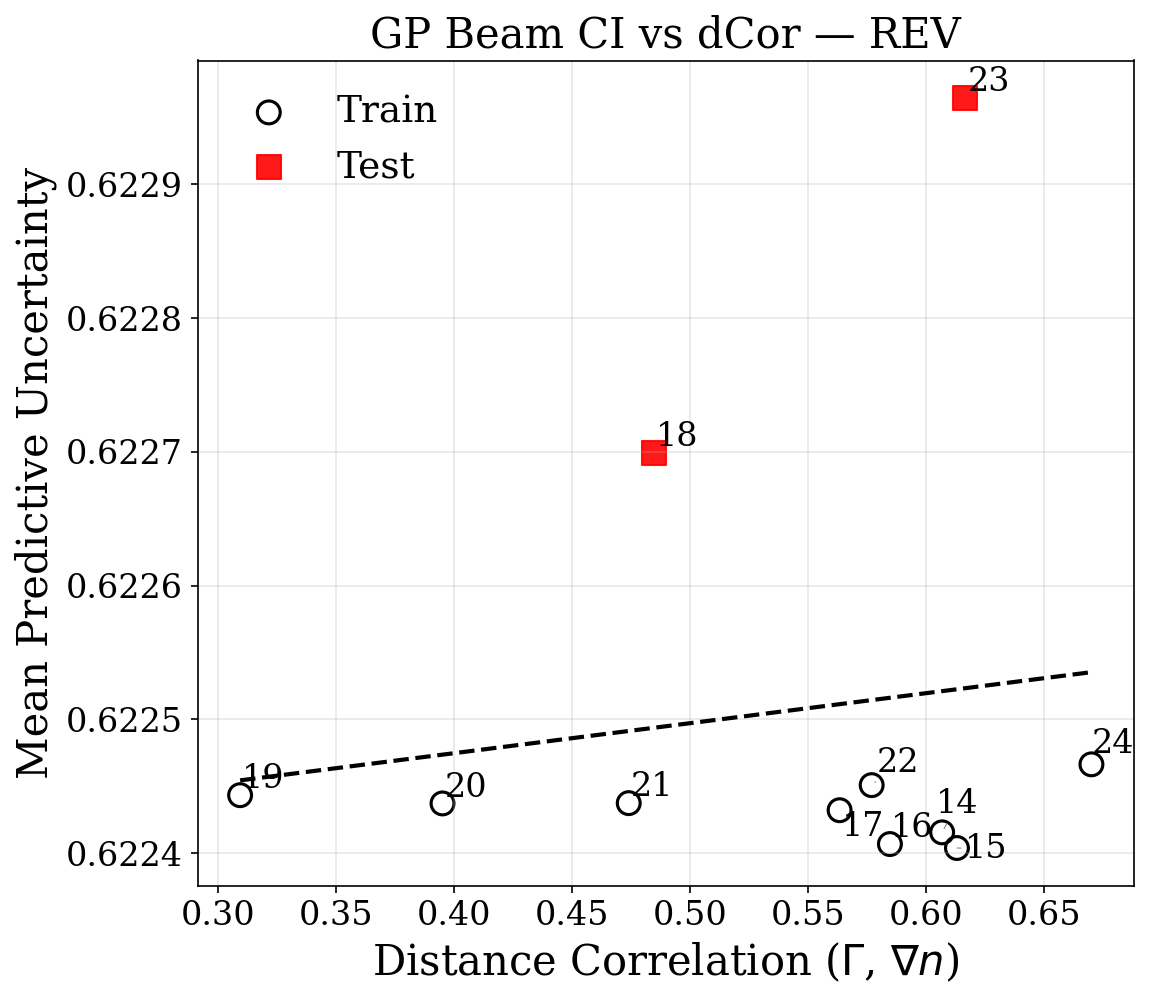}
        \subcaption{}
        \label{fig:subfig1}
    \end{subfigure}
    \hfill
    \begin{subfigure}{0.48\textwidth}
        \centering
        \includegraphics[width=\textwidth]{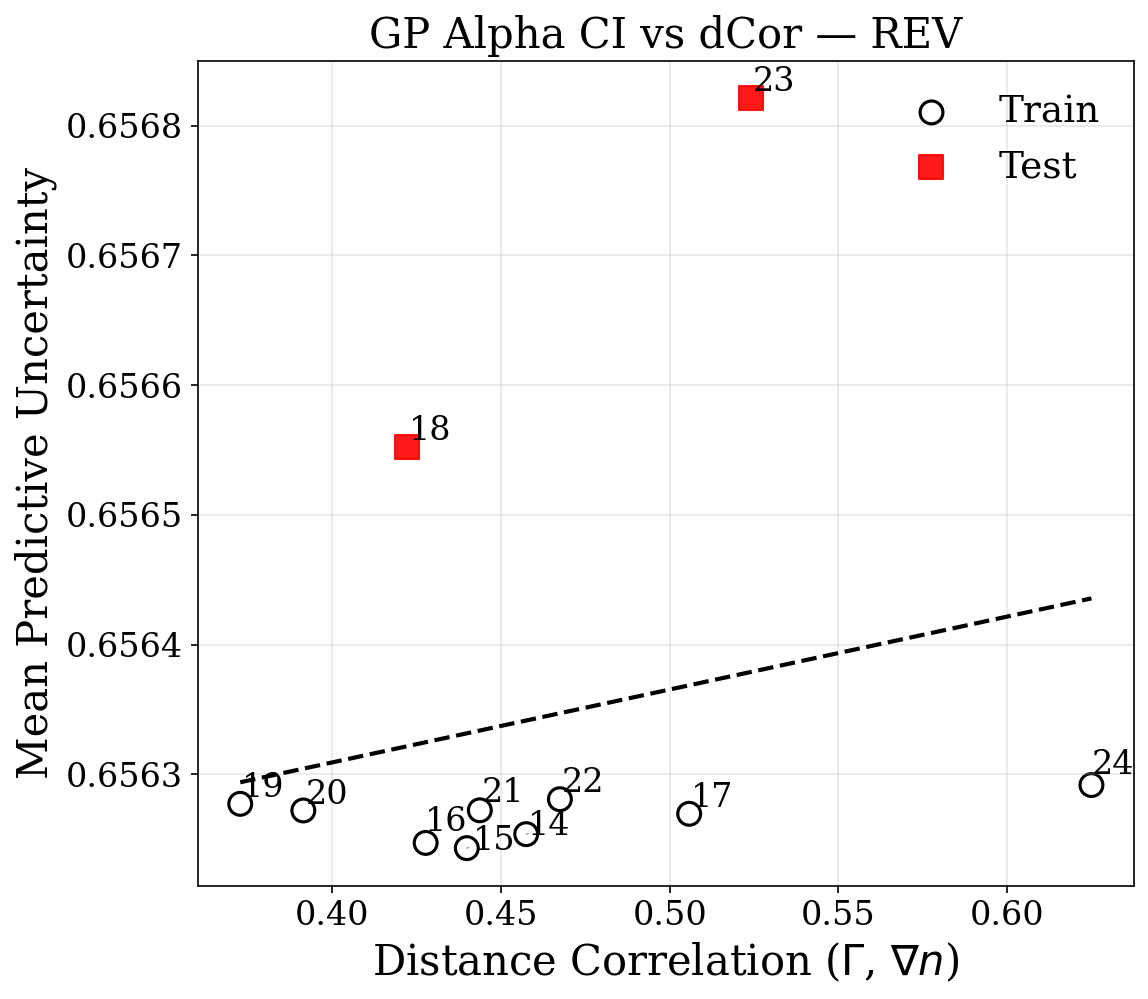}
        \subcaption{}
        \label{fig:subfig2}
    \end{subfigure}

    \vspace{0.5cm}

    \begin{subfigure}{0.48\textwidth}
        \centering
        \includegraphics[width=\textwidth]{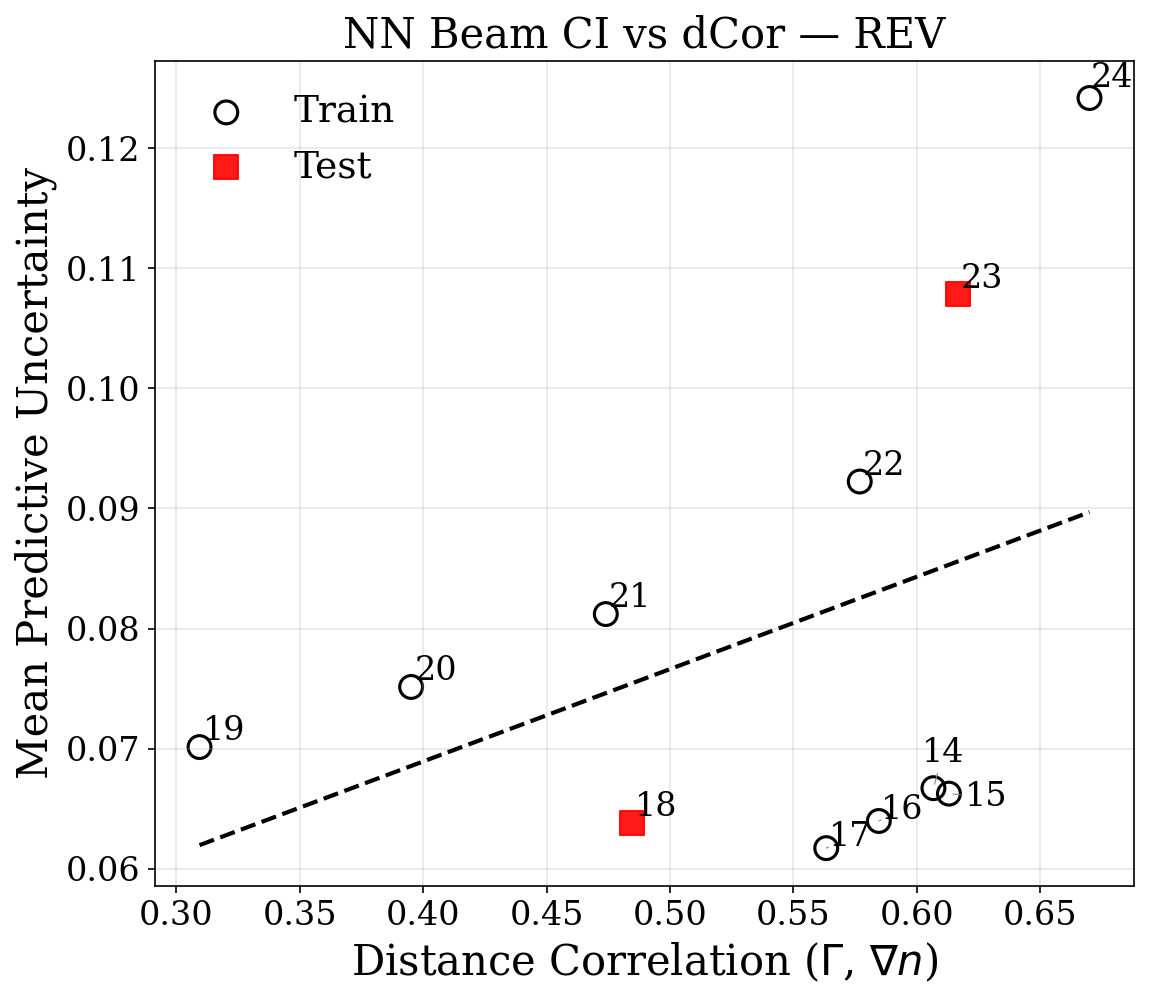}
        \subcaption{}
        \label{fig:subfig3}
    \end{subfigure}
    \hfill
    \begin{subfigure}{0.48\textwidth}
        \centering
        \includegraphics[width=\textwidth]{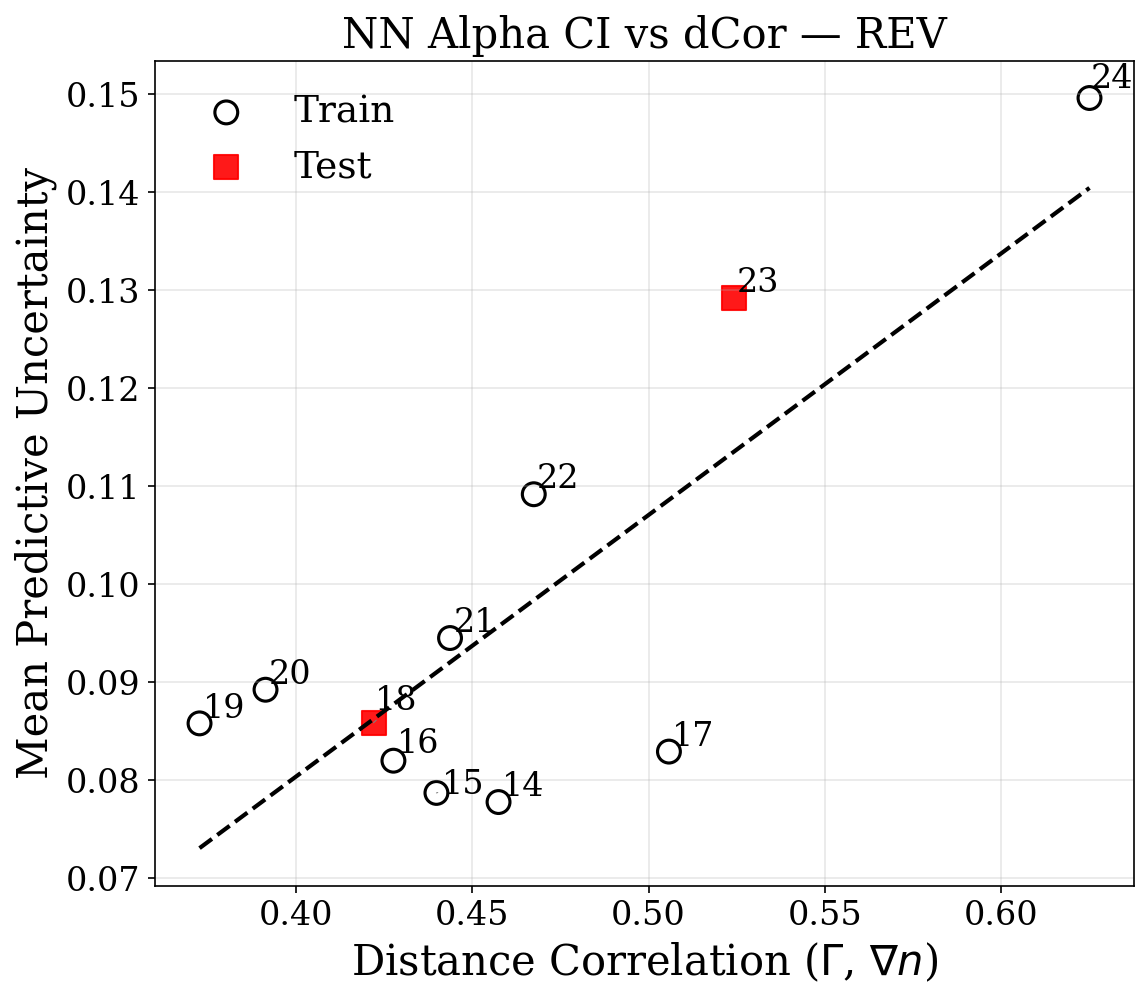}
        \subcaption{}
        \label{fig:subfig4}
    \end{subfigure}

    \caption{Plots of mean predictive uncertainty in flux profile calculation for ITER steady state case with reversed shear q-profile with respect to distance correlation between EP fluxes and corresponding EP density gradients for energetic beam (figs. (a) and (c)) and alpha particles (figs. (b) and (d)). Profiles labeled 14-18 correspond to the early and profiles labeled 19-24 correspond to the late saturation phase in the nonlinear FAR3d simulations for the ITER reversed-shear case.}
    \label{fig:CIvsdcor}
\end{figure}

\paragraph{Correlation between predictive uncertainty and transport regime:} It is noteworthy that the early saturated phase of the simulations is characterized by a stronger distance correlation coefficient between the EP transport fluxes and the EP density gradients, as is also expected from a physics point of view. As the simulations proceed towards a regime later in time, the distance correlation between EP fluxes and density gradients drops to a lower value at the transition step (denoted by output dump file 19) and then increase gradually  to large values of distance correlation as the simulation progresses. In Figure \ref{fig:CIvsdcor}, we examine the relationship between mean uncertainties per profile in the EP flux profile predictions as calculated by GP and NN models and the correlation between fluxes and EP gradients for the ITER reversed shear case. As shown in Fig. \ref{fig:CIvsdcor}, the GP surrogate model assigns a near-uniform predictive uncertainty to all transport flux profiles, whereas the NN model has more variation in the uncertainty width predictions for different profiles. These results also explain the relatively narrow distribution of profile-wise $L^2$ errors observed for the GP model in contrast to the NN model that exhibits a stronger variation in predictive uncertainty with transport regime and a broader distribution of relative $L^2$ errors. The NN model exhibits a clear separation in uncertainty levels between the early and late saturation regimes, with the uncertainties in these two regimes established on the two opposite sides of the diagonal line. This behavior suggests that the NN model is more sensitive to changes in the underlying transport regime for evaluation of EP transport fluxes. 
\subsection{Representative Transport Profiles}
\begin{figure*}[htb!]
\centering
\captionsetup[subfigure]{labelfont=small,textfont=small}
{\textbf{Gaussian process (GP)}}

\vspace{0.15cm}


\subfloat[]{
    \includegraphics[width=0.25\textwidth]{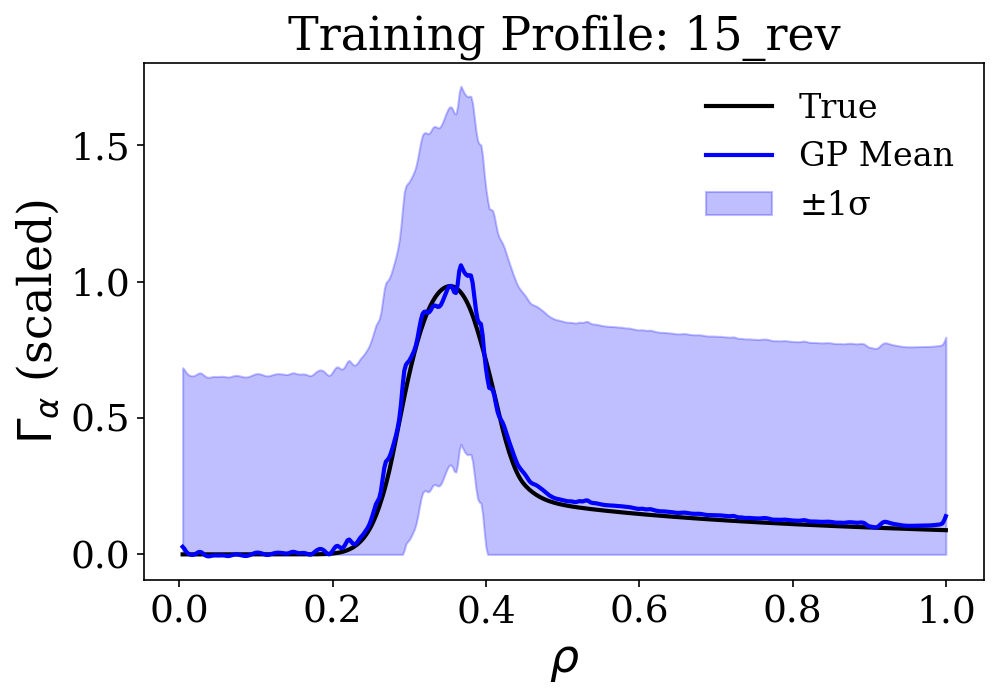}
}
\subfloat[]{
    \includegraphics[width=0.25\textwidth]{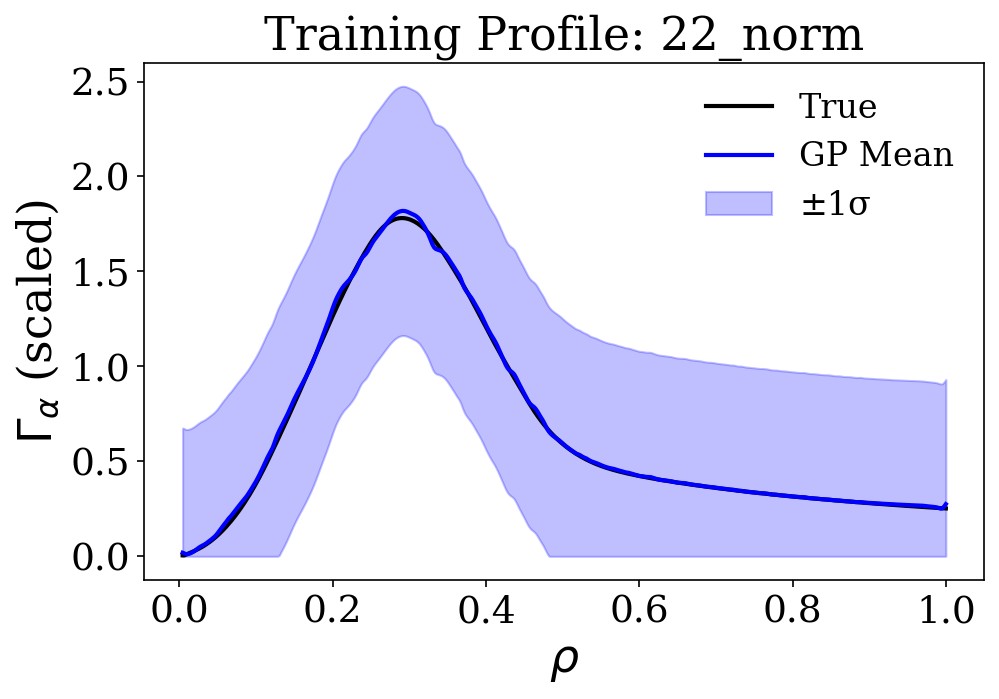}
}
\subfloat[]{
    \includegraphics[width=0.25\textwidth]{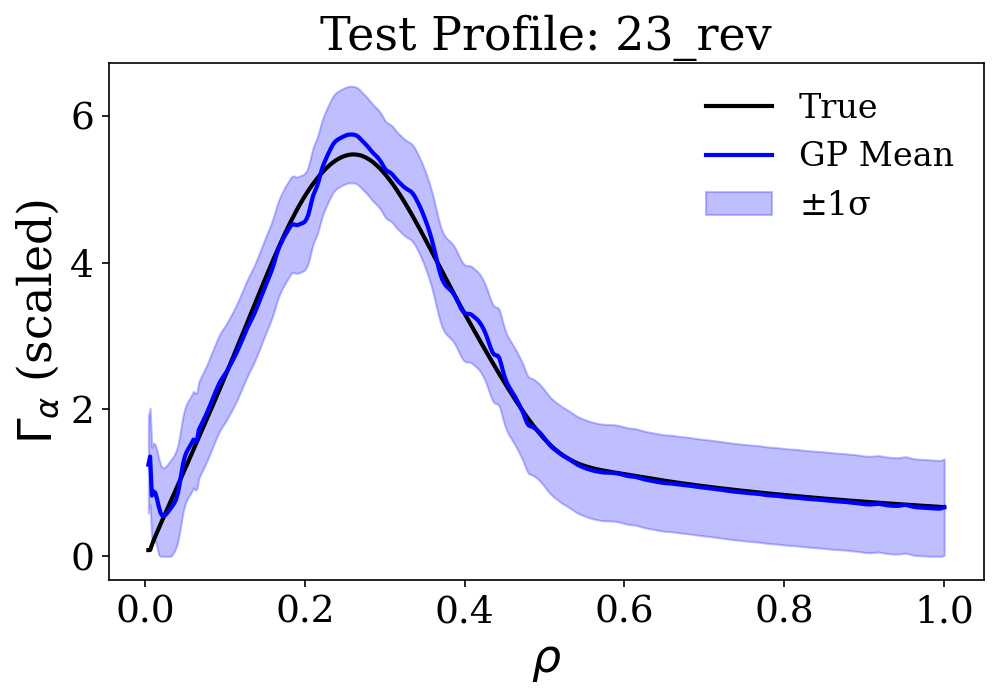}
}
\subfloat[]{
    \includegraphics[width=0.25\textwidth]{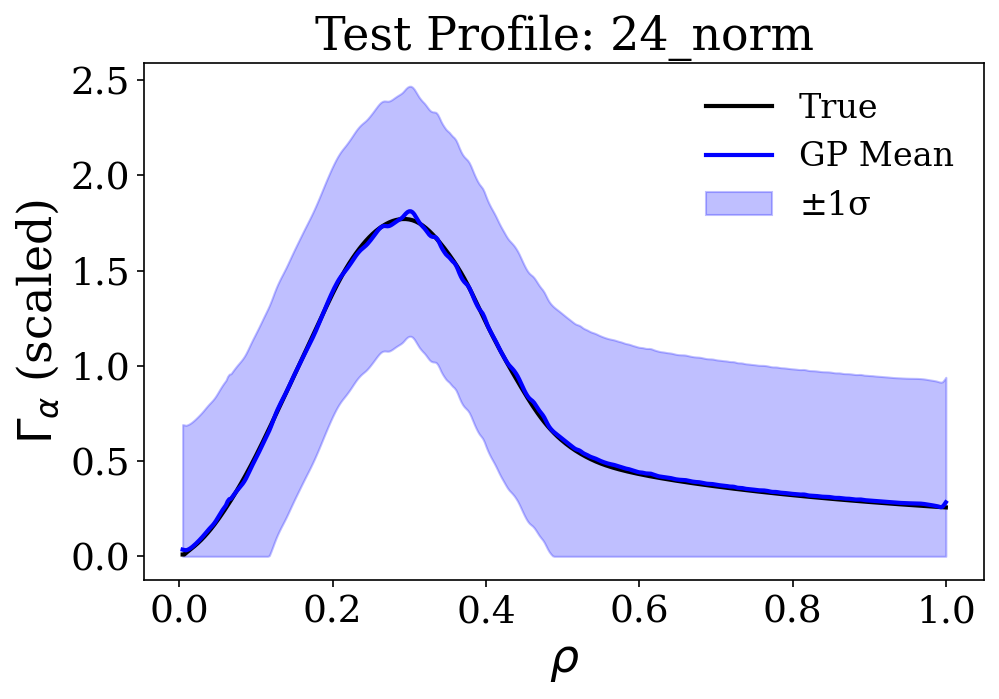}
}

\vspace{0.3cm}

{\textbf{Neural Network (NN)}}

\vspace{0.15cm}


\subfloat[]{
    \includegraphics[width=0.25\textwidth]{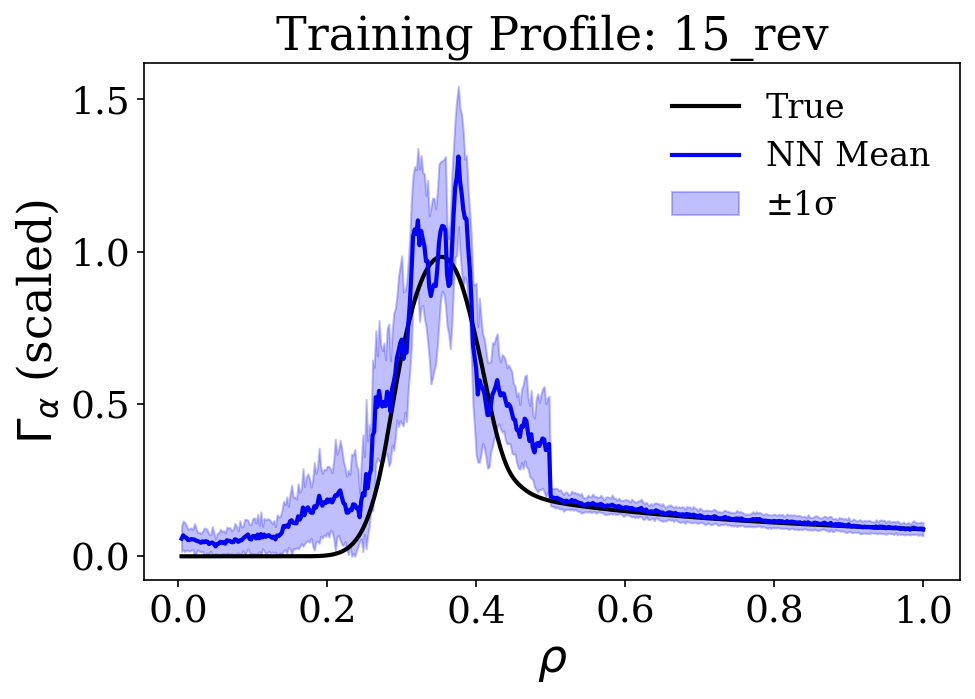}
}
\subfloat[]{
    \includegraphics[width=0.25\textwidth]{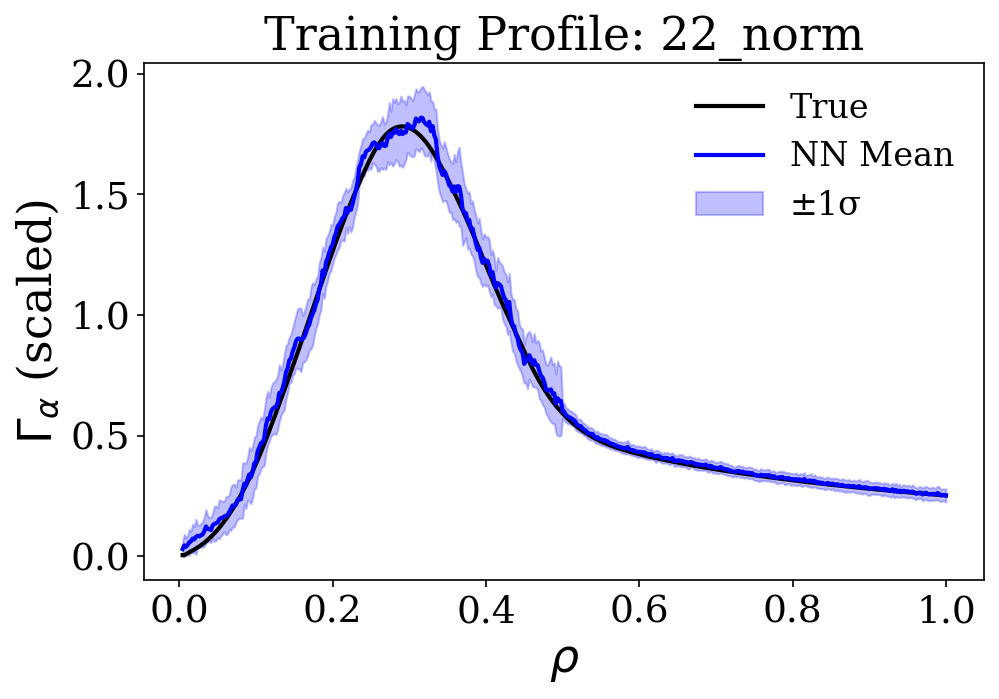}
}
\subfloat[]{
    \includegraphics[width=0.25\textwidth]{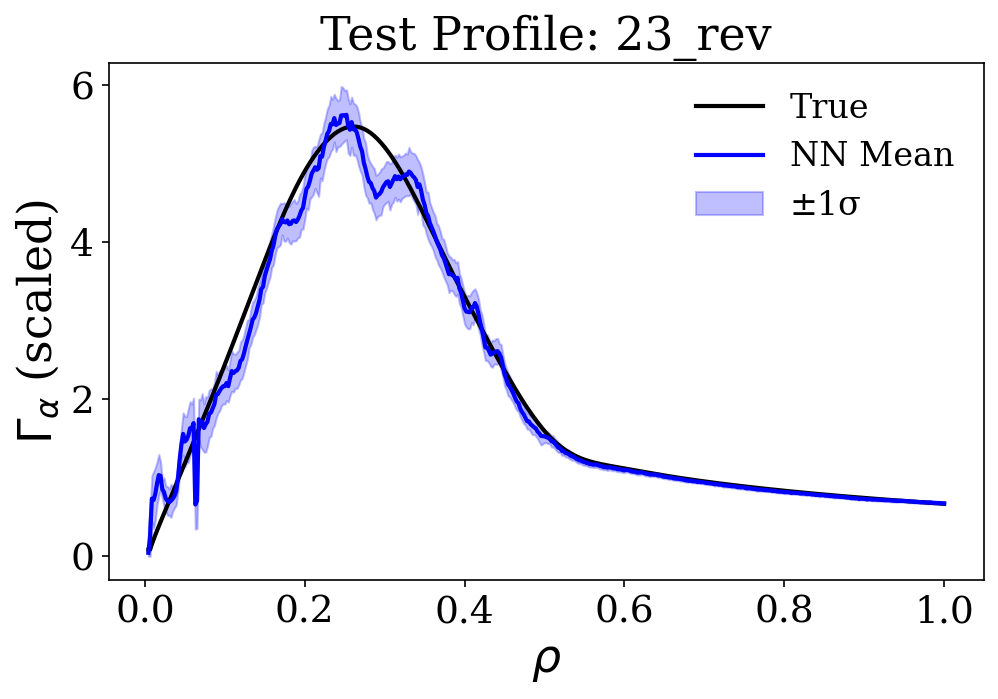}
}
\subfloat[]{
    \includegraphics[width=0.25\textwidth]{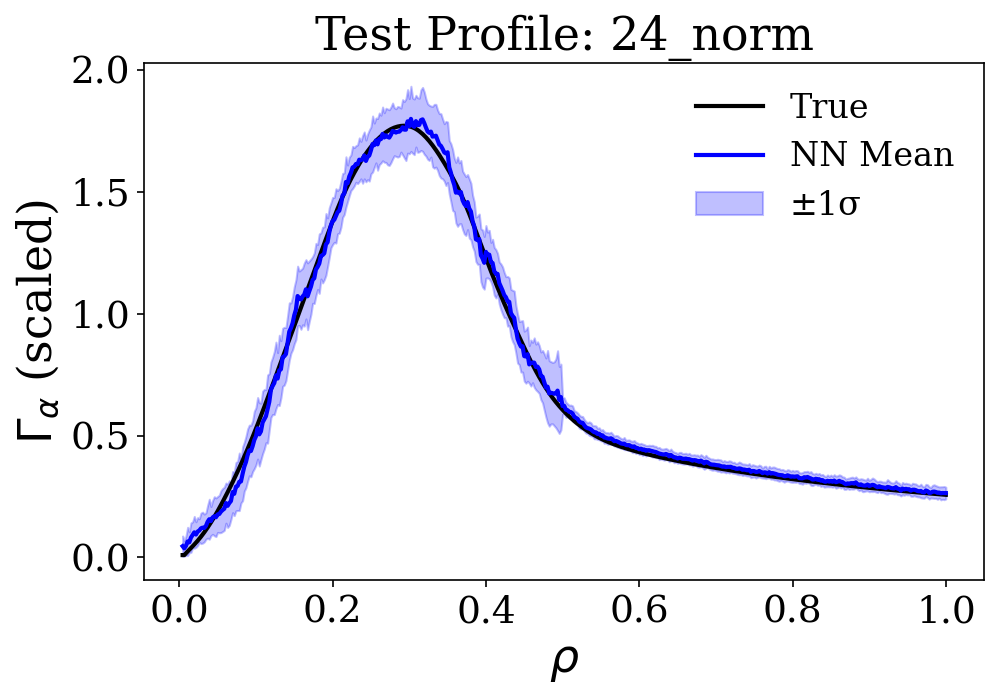}
}

\caption{Representative alpha-particle transport flux profiles predicted by the GP (top row) and NN (bottom row) surrogate models. The selected profiles include representative training and testing cases for the reversed-shear and monotonic-$q$ ITER configurations, illustrating both challenging and well-predicted transport profiles. Shaded regions denote $\pm 1\sigma$ predictive uncertainty, obtained from the GP predictive distribution and MC-dropout realizations for the GP and NN surrogates, respectively. The transport fluxes are shown in normalized units as defined in Sec.~\ref{section:radial_discretization}.}
\label{fig:profile_comparison}

\end{figure*}

To further illustrate the profile-wise reconstruction quality discussed above, Fig. \ref{fig:profile_comparison} compares representative transport flux profiles predicted by the GP and NN surrogate models. Figures \ref{fig:profile_comparison}(a), (c), (e), and (g) correspond to representative higher-error profiles from the reversed-shear ITER case identified in the error analysis in Fig. \ref{fig:l2_boxplots}, whereas Figs. \ref{fig:profile_comparison} (b), (d), (f), and (h) show representative flux profiles for the monotonic-$q$ ITER case that are accurately reconstructed by both surrogate models. Although the prediction intervals represented by one standard deviation are generally narrower for the NN surrogate, the uncertainty estimates from the GP and NN models have different interpretations, as discussed in the previous section \ref{subsection:uncertainty_quantification}. For the representative profiles shown here, the GP mean predictions generally remain closer to the true transport profiles than those of the NN model. It is also evident that both surrogate models reconstruct the EP transport fluxes for the monotonic-$q$ profiles more accurately than the reversed-shear profiles. This behavior is likely associated with the stronger nonlinear transport dynamics in the reversed-shear case, which make the dependence of the transport fluxes on the EP density gradients more difficult to learn.
\par
Overall, the GP and NN surrogate models achieve comparable predictive accuracy for both beam and alpha-particle transport fluxes. The GP model generally provides slightly more accurate mean profile reconstructions, whereas the NN model exhibits predictive uncertainties that more clearly distinguish between different transport regimes. These complementary characteristics make both surrogate approaches attractive for reduced energetic-particle transport modeling.
\section{Conclusions and future work directions}

In this work, we presented, to our knowledge, the first machine-learning surrogate models for predicting Alfv\'en eigenmode-driven energetic-particle transport fluxes in an ITER steady-state scenario. By training directly on nonlinear FAR3d simulations, the surrogate models retain the nonlinear transport response of the underlying physics while reducing the computational cost of transport evaluation by approximately five to six orders of magnitude relative to direct nonlinear simulations. Such a reduction makes these surrogate models attractive for future incorporation into integrated modeling workflows requiring repeated evaluations of energetic-particle transport.

\par

A flux-consistency analysis demonstrated that the selected seven-dimensional plasma-state representation provides a sufficiently unique parameterization of the nonlinear transport response over most of the sampled feature space, thereby justifying the surrogate formulation. Two complementary surrogate methodologies, a multitask Gaussian process (GP) regression model and a hierarchical neural-network (NN) model, were developed and systematically compared. Both surrogate models achieved comparable global prediction accuracy, while the GP model generally produced more accurate reconstruction of the radial transport flux profiles, particularly in regions of strong energetic-particle transport and large energetic-particle density gradients.

\par

The two surrogate approaches exhibited distinct uncertainty characteristics. The GP naturally provides predictive uncertainty through its probabilistic formulation and produced relatively smaller profile-to-profile variations in predictive uncertainty throughout the dataset. In contrast, uncertainty estimates obtained using the Monte Carlo dropout approach showed that the NN surrogate was more sensitive to variations in the underlying transport regime, distinguishing more clearly between the early and late stages of nonlinear saturation. These complementary characteristics suggest that GP and NN surrogate models offer different advantages depending on the intended application.

\par

The surrogate developed in this work is intended to replace repeated evaluations of the nonlinear energetic-particle transport response within integrated transport calculations, rather than reproducing the full temporal evolution of Alfv\'en eigenmode growth. Since energetic-particle redistribution becomes significant only after nonlinear saturation has been established, the surrogate was trained exclusively using transport data from the nonlinear saturated phase of the simulations. The applicability of the present surrogate is limited to the nonlinear saturated transport regimes represented in the training database; its applicability to strongly driven bursting or large-scale relaxation regimes remains to be assessed. 
\par Future work will extend the present methodology in two directions. First, surrogate models will be developed for present-day experimental devices, such as DIII-D, where nonlinear simulation datasets can be generated more readily and validated against experimental observations. Second, we are developing surrogate models capable of predicting the time-dependent evolution of energetic-particle transport, with the long-term goal of accelerating nonlinear simulations themselves and generating larger synthetic datasets for improving surrogate-model accuracy. The present surrogate was trained for a single ITER steady-state scenario; extension to multiple operating scenarios will require additional nonlinear training data and constitutes an important direction for future work. These developments will further extend the applicability of machine-learning surrogate models for predictive energetic-particle transport modeling in future fusion reactors.
\section*{Acknowledgments}

This research was sponsored by the Laboratory Directed Research and Development Program of Oak Ridge National Laboratory, managed by UT-Battelle, LLC, for the U.S. Department of Energy, with additional support through the MiRACL FIRE Collaborative. The authors thank Aditya Kashi, Sebastian de Pascuale and Mark Cianciosa of Oak Ridge National Laboratory and Arpan Biswas of the University of Tennessee, Knoxville, for helpful discussions on machine-learning methods for surrogate-model development.

\bibliographystyle{unsrt}
\bibliography{sample}

\end{document}